\documentclass[useAMS,usenatbib,onecolumn]{mnras}

\usepackage{dcolumn}
\usepackage{graphicx}
\usepackage{url}
\usepackage{color}
\usepackage{mathtools}

\usepackage{longtable}

\usepackage{booktabs}

\newcommand{\cm}{cm$^{-1}$}

\newcommand{\ai}{\textit{ab initio}}
\newcommand{\Ai}{\textit{Ab initio}}

\newcommand{\duo}{{\sc Duo}}

\newcommand{\X}{$X$~$^2\Pi$}

\newcommand{\A}{$A$~$^{2}\Sigma^{+}$}

\newcommand{\ket}[1]{\vert #1 \rangle  }
\newcommand{\bra}[1]{\langle #1 \vert  }

\title[ExoMol XXVI: Line list for SH]{ExoMol molecular line lists  - XXVI:  spectra of SH and NS}

\date{\today}
\author[Yurchenko et al.]{Sergei N. Yurchenko$^{1}$, Wesley Bond$^{1}$, Maire N. Gorman$^{1,2}$,  Lorenzo Lodi$^{1}$,\newauthor
Laura K. McKemmish$^{1}$,
 William Nunn$^{1}$, Rohan Shah$^{1}$ and Jonathan Tennyson$^{1}$\thanks{Email: j.tennyson@ucl.ac.uk}  \\
$^{1}$ Department of Physics and Astronomy, University College London, London WC1E 6BT,
UK\\
$^{2}$ Department of Physics, Aberystwyth University, Penglais, Aberystwyth, Ceredigion, UK, SY23 3BZ}
\date{Accepted XXXX. Received XXXX; in original form XXXX}
\pubyear{2018}
\pagerange{\pageref{firstpage}--\pageref{lastpage}}
\begin{document}
\maketitle

\begin{abstract}

Line lists for the sulphur-containing molecules SH (the mercapto
radical) and NS are computed as part of the ExoMol project. These
line lists consider transitions within the $X$~${}^2\Pi$ ground state
  for $^{\text{32}}$SH, $^{\text{33}}$SH, $^{\text{34}}$SH and
  $^{\text{32}}$SD,  and
  $^{14}$N$^{32}$S, $^{14}$N$^{33}$S, $^{14}$N$^{34}$S,
  $^{14}$N$^{36}$S and $^{15}$N$^{32}$S.  \Ai\ potential energy (PEC)
 and spin-orbit coupling (SOC) curves
are computed and then improved by fitting to experimentally
  observed transitions. Fully \textit{ab initio} dipole moment curves
  (DMCs) computed at high level of theory are used to produce the final line lists.
  For SH, our fit gives
  a root-mean-square (rms) error of 0.03 cm$^{-1}$ between the
  observed ($v_{\rm max}=4$, $J_{\rm max} = 34.5$) and calculated
  transitions wavenumbers; this is
  extrapolated such that all \X\ rotational-vibrational-electronic (rovibronic) bound states
  are considered. For   $^{\text{32}}$SH the resulting line list contains about 81~000 transitions and 2~300 rovibronic states, considering levels up to   $v_{\rm max} = 14$ and $J_{\rm max} = 60.5$.  For NS the refinement used a combination of experimentally determined frequencies and energy levels and led to an rms fitting error  of 0.002~\cm. 
Each NS calculated line list includes around 2.8 million transitions and 31~000  rovibronic states  with a vibrational range up to $v=53$ and rotational range
  to $J=235.5$, which covers up to 23~000 \cm. Both line lists should be complete for temperatures up to 5000~K. Example  spectra simulated using this line list are shown and comparisons
  made to the existing data in the CDMS database.  The line lists are available from the CDS
  (http://cdsarc.u-strasbg.fr) and ExoMol (www.exomol.com) data bases.
\end{abstract}


\section{Introduction}


\label{firstpage}

Sulphur chemistry is important in a variety of astronomical
environments including the interstellar medium (ISM)
\citep{74OpDaXX.NS,80DuMiWi.SH,17ViLoJa.SH}, hot cores
\citep{97Charnl.NS,15WoOcVi.NS}, comets
\citep{02CaAlBo.NS,06RoChXX.NS,07CaAlBo.NS}, starburst and other
galaxies \citep{05MaMaMa.NS,05Martin.NS}, exoplanets
\citep{06ViLoFe.SH,09ZaMaFr.SH}, and brown dwarfs and low-Mass dwarf
stars \citep{06ViLoFe.SH}. The ExoMol project aims at providing
comprehensive molecular line lists for exoplanet and other
atmospheres. ExoMol has provided line lists for several sulphur-baring
molecules: CS \citep{jt615}, SO$_2$ \citep{jt635}, H$_2$S
\citep{jt640}, SO$_3$ \citep{jt641}, and PS \citep{jt703}; a line list
for SiS has also just been completed \citep{jt724}. In this
work we extend this coverage by providing line lists for the major
isotopologues of SH and NS. For both species we only consider
transitions within the ground electronic state: both SH and NS have an
\X\ ground state. The excited electronic  states lie above 30,000 \cm{} and 23,000~\cm\ for SH and NS, respectively, and thus the line lists presented here will be accurate for the visible, infrared and radio spectral regions.   Both species are well-know astronomically from transitions within the ground state.

The diatomic mercapto radical SH has long been of interest to astronomers, but proved challenging to detect. It was definitively detected
in the interstellar medium (ISM) \citep{12NeFaGe.SH}, in asymptotic-giant-branch (AGB) stars \citep{00YaKaRi.SH} and the Sun's atmosphere \citep{02BeLiXX.SH},
tentatively detected in comets \citep{87SwWaXX.SH,88KrWaXX.SH} and predicted to occur in brown dwarfs \citep{06ViLoFe.SH}
and hot Jupiter exoplanets \citep{06ViLoFe.SH,09ZaMaFr.SH} as one of the major sulphur-bearing gases after H$_2$S.
The ISM detection was difficult due to the location of the key rotational transition which was inaccessible both from the ground
and the Herschel telescope;
after a number of failed searches in the ISM
\citep{69MeGoLi.SH,71HeTuXX.SH}, \citet{12NeFaGe.SH} finally detected
SH in the terahertz region using SOFIA (Stratospheric
Observatory For Infrared Astronomy) by its 1383 GHz $^{2}\Pi_{\frac{3}{2}}\
  J=\frac{5}{2} - \frac{3}{2}$ transition.
  The \A--\X{} band, a UV absorption band not considered in this paper, has also been used to detect SH
  in translucent interstellar clouds \citep{15ZhGaLi.SH} and in the Sun's atmosphere  \citep{02BeLiXX.SH}.
 \citet{09ZaMaFr.SH} generated an $A$--$X$ line list for SH.

In our own atmosphere, SH is known to react with NO$_{2}$, O$_{2}$ and O$_3$: SH is produced in the troposphere by oxidation
of H$_2$S by the OH radical \citep{94RaWiFl.SH}.

Experimentally, SH spectra have been studied since 1939 \citep{39GlHoXX.SH,39LeWhXX.SH} with over ~100 experimental publications to date.
 This work is based on the measured transitions from experimental studies presented in Table~\ref{t:HSlit}.

A number of multi-reference configuration interaction (MRCI) level
theoretical calculations have been performed on SH
\citep{75RaTaSi.SH,82HiGuxx.SH,87BrHixx.SH}, with the most recent study
been those of \citet{17KaTaHi.SH} and \citet{16VsNsxx.SH}. Spin-orbit
splitting of the ground state potential energy curve (PEC)
was calculated by \citet{90BaLexx.SH} and
\citet{08LiGaZh.SH}. Lifetimes for SH have previously been
calculated by \citet{98Mcxxxx.SH} and \citet{08BrHaHo.SH} with
\citet{01ReOrx1.SH}.


Transitions for NS are much more astronomically accessible
and it was one of the first ten diatomic molecules to be detected in space \citep{77Somerv.NS, 79LoJoSn.NS},
with the first positive detection by \citet{75GoBaGo.NS} using the $J=5/2 - 3/2$ transition
of the $^2\Pi_{1/2}$ state at 115.6 GHz towards Sagittarius B2. Radio astronomy has
also been used to detect NS in giant molecular clouds \citep{92McIrMi.NS,06LeRoTh.NS,13BeMuMe.NS},  cold dark clouds \citep{94McIrOh.NS}, comets \citep{99IrLoSe.NS,05Biver.NS}, extragalactically \citep{03MaMaMa.NS} and the NGC 253 starburst region \cite{15MeWaBo.NS}.

Experimentally, there have been significant experimental work focusing on the excited electronic states;
however, this is not of relevance to this work.
Numerous experimental studies on NS
have been made on the spectra of the ground state. Laser magnetic resonance (LMR) studies
include those of \citet{68CaHoLe.NS}, \citet{69UeMoXX.NS}, \citet{94Anacon.NS}
and \citet{95Anacon.NS}. Experimental measurements of rovibrational transitions
within the ground state are reported in a series of papers \citep{71NaBaXX.NS,80MaKaNa.NS,82LoSuXX.NS,86AnBoDa.NS,88SiBuHa.NS,95LeOzSa.NS,69AmSaHi.NS}.
The experimental frequencies used in this work are summarised in Table~\ref{t:NSlit}.

Early electronic structure calculation on NS were made by
\citet{76BiGrXX.NS}, \citet{76SaMeXX.NS} and \citet{78KaScPe.NS}.
  Subsequently, \citet{85LiPeBu.NS} and \citet{86KaGrXX.NS}
performed configuration interaction studies on the low-lying and
Rydberg states of NS. CCSD(T) calculation of equilibrium geometries of
NS for plasma applications were made by \citet{04CzViXX.NS}. The most
recent theoretical study on NS is that of \citet{13GaGaGo.NS} who
undertook calculations at the MRCI+Q+DK/AV5Z level of theory
for the PECs of low-lying electronic states.
However, while there are some computed dipole moment \citep{85LiPeBu.NS,13GaGaGo.NS}
and spin-orbit coupling \citep{12ShXiSu.NS} curves,
we perform new \ai\ calculations to ensure the uniform quality of our model.

\section{Method and Spectroscopic Models}

Our general method is to start from high quality \ai\ potential energy
curves (PECs), associated coupling curves  and dipole moment curves
(DMCs). Since \ai\ transition frequencies are not accurate enough, the
PECs and couplings are refined using empirical energy levels and transition wavenumbers from laboratory spectra.   \Ai\ DMCs are found to give the best results. The
nuclear motion problem is solved using the program Duo \citep{jt609}
which allows for full couplings between the curves, see \citet{jt632}
for a full discussion of the theory. \duo\ has been successfully used
to produce line lists for a number of diatomic molecules AlO, PS, PN,
ScH, VO, NO, CaO, SiH
\citep{jt598,jt599,jt644,jt618,jt686,jt703,jt711}. The refined PECs, coupling curves and DMCs together form a spectroscopic model for the diatomic system, which can be useful beyond the immediate line list application considered here.

Both SH and NS have \X\ ground states. In this case the
spin-orbit (SO) coupling splits the PEC in two curves, which are often
denoted $^{2}\Pi_\frac{1}{2}$ and $^{2}\Pi_\frac{3}{2}$. The SO
splitting coupling presents a significant contributions to the
energies of these molecules: 360~\cm\ and 220~\cm\ for the $v=0$ states of SH and NS, respectively. Another important coupling for
spectroscopy of the $^{2}\Pi$ systems is the due to the presence of electronic angular
momentum (EAM), which causes the $\Lambda$-doubling effect.

We use the extended Morse oscillator (EMO) functions \citep{EMO} to represent the PECs, both \ai\ and refined.
In this case the PEC is given by
\begin{equation}\label{e:EMO}
V(r)=V_{\rm e}\;\;+\;\;(A_{\rm e} - V_{\rm
e})\left[1\;\;-\;\;\exp\left(-\sum_{k=0}^{N} B_{k}\xi_p^{k}(r-r_{\rm e})
\right)\right]^2,
\end{equation}
where $A_{\rm e} - V_{\rm e}$ is the dissociation energy, $r_{\rm e}$ is an equilibrium distance of the PEC, and $\xi_p$ is the \v{S}urkus variable given by:
\begin{equation}
\label{e:surkus:2}
\xi_p= \frac{r^{p}-r^{p}_{\rm e}}{r^{p}+r^{p}_{\rm e }}.
\end{equation}
The corresponding expansion parameters are obtained by fitting to the experimental data (energies and frequencies) of the molecule in question, as detailed below.

To model the SO coupling we use \ai\ curves computed using high levels
of theory with the program MOLPRO \citep{MOLPRO}. These curves are
then refined by fitting to the experimental data using the morphing
approach \citep{99MeHuxx.methods,99SkPeBo.methods}. In this approach, the \ai\ curves
represented on a grid of bond lengths are `morphed' using the
following expansion:
\begin{equation}
\label{e:bob}
F(r)=\sum^{N}_{k=0}B_{k}\, z^{k} (1-\xi_p) + \xi_p\, B_{\infty},
\end{equation}
where $z$ is either taken as the \v{S}urkus variable $z=\xi_p$  or the
damped-coordinate given by:
\begin{equation}\label{e:damp}
z = (r-r_{\rm ref})\, e^{-\beta_2 (r-r_{\rm ref})^2-\beta_4 (r - r_{\rm ref})^4},
\end{equation}
see also \citet{jt703} and \citet{jt711}.
Here $r_{\rm ref}$ is a reference position equal to $r_{\rm e}$ by default and $\beta_2$ and $\beta_4$ are damping factors. When used for morphing, the parameter $B_{\infty}$ is usually fixed to 1.

The  $\Lambda$-doubling effects in \duo\ can be modelled directly using an effective $\Lambda$-doubling function, in case of $^2\Pi$ we use the $(p+2q)$ effective coupling \citep{79BrMexx.methods} given by:
\begin{equation} \label{e:Hlambda-doubling}
\hat{H}_{\rm LD} =
- \frac{1}{2} \alpha_{p2q}^{\rm LD }(r)  \left( \hat{J}_{+}\hat{S}_{+} + \hat{J}_{-}\hat{S}_{-} \right).
\end{equation}
$\hat{H}_{\rm LD}$ leads to a linear $\hat{J}$-dependence, which is justified for the heavy  molecule like NS. In this case for  $\alpha_{p2q}^{\rm LD }(r)$ we use a simple, one-parameter function:
\begin{equation} \label{e:ap2q}
\alpha_{p2q}^{\rm LD } = B_{0}^{p2q}\, (1-\xi_p).
\end{equation}
For the SH molecule, which is affected by a stronger
centrifugal distortion, this is not appropriate.  Here we follow the
approach recently used for solving another hydrogen-containing $^2\Pi$ system,
SiH \citep{jt711}, where the $\Lambda$-doubling is
modelled via an EAM interaction with a closely lying $^2\Sigma$-state
\citep{79BrMexx.methods}. In the case of \X\ of SH, the
closest $\Sigma$ state is  $A$~$^2\Sigma^-$.  The latter is introduced
with a dummy potential curved in the EMO representation, while the
EAM-curve is given by the 1st order $\xi_p$-type expansion in
Eq.~\eqref{e:bob} (see also below).

The dipole moment curves (DMC) of SH and NS are computed using a high level \ai\
theory on a grid of bond length values ranging from about 0.8 to 8~\AA. In order
to reduce the numerical noise in the intensity calculations of high overtones
(see recent recommendations by \citet{16MeMeSt} the DMCs are represented
analytically. The expansion with a damped $z$ coordinate in Eq.~\eqref{e:bob} is
employed \citep{jt703,jt711}.

All these functional forms are included in \duo\ (functions.f90). The
corresponding expansion parameters as well as their grid representations
can be found in the \duo\ input files provided as supplementary data.

\subsection{SH}

\begin{figure}
\centering
\includegraphics[width=0.49\textwidth]{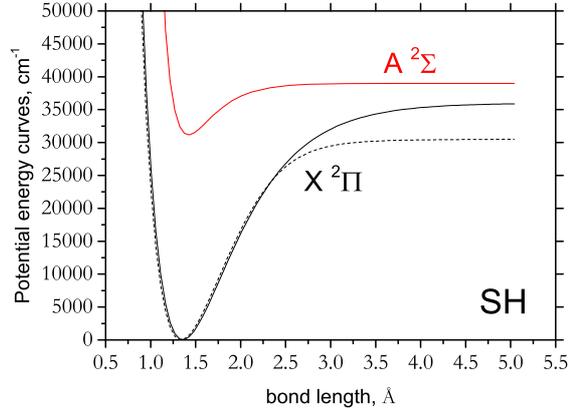}
\caption{Potential energy curves of SH: fitted (solid) and \ai\ (dashed).}
\label{f:SH:PECs}
\end{figure}

\begin{figure}
\centering
\includegraphics[width=0.49\textwidth]{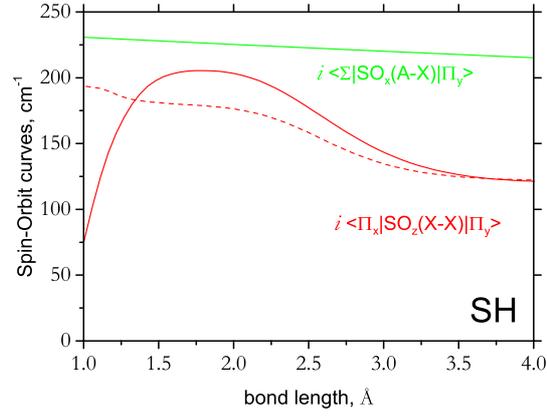}
\caption{Spin-orbit curves of SH: $X$--$X$,  fitted (solid) and \ai\ (dashed) and $A$--$X$, fitted only (solid).}
\label{f:SH:SO}
\end{figure}

\begin{figure}
\centering
\includegraphics[width=0.49\textwidth]{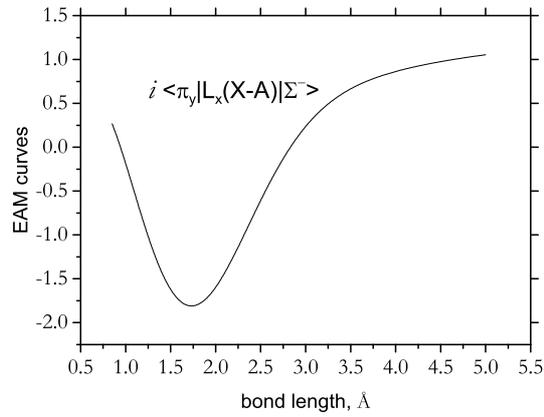}
\caption{An empirical EAM curve ($A$--$X$) of SH.}
\label{f:SH:Lx}
\end{figure}

\begin{figure}
\centering
\includegraphics[width=0.49\textwidth]{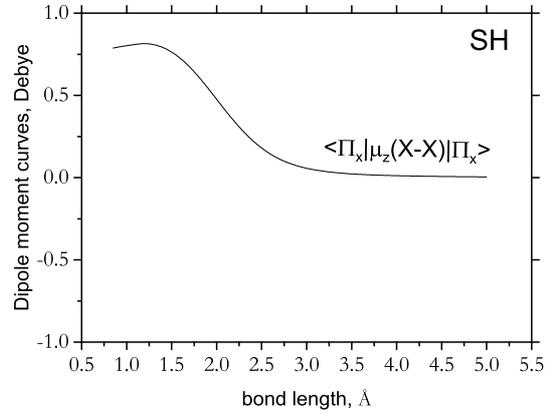}
\caption{The diagonal  \X\ \ai\ dipole moment curve of SH.}
\label{f:SH:DMC}
\end{figure}

The \duo\ model for SH consists of two PECs, \X\ and $A$~$^{2}\Sigma^{+}$, represented by EMO forms in Eq.~\eqref{e:EMO},
diagonal ($X$--$X$) and non-diagonal ($X$--$A$) SO coupling curves
(\ai) morphed by functions using Eq.~\eqref{e:bob}, an EAM coupling
curve ($X$--$A$) also represented by Eq.~\eqref{e:bob}, an by a diagonal
$X$--$X$ DMC.  The $A$ state PEC is only used to support the
$\Lambda$-doubling effect in the $X$-state energies and is not
included in SH the line list. We use the \ai\ SO coupling curves
obtained at the MRCI+DKH4+Q level of theory, where DKH4 is the fourth-order
Douglas-Kroll-Hess representation of the relativistic Hamiltonian
and Q denotes a Davidson correction. An AWC5Z Gaussian Type basis set
was used
\citep{89Dunnin.ai,93WoDuxx.ai,02PeDuxx.ai,12SzMuGi.ai}.
The \ai\ PE, SO, EAM and DM curves of SH used in this work are shown in Figs~\ref{f:SH:PECs}--\ref{f:SH:DMC}.



\begin{figure}
\centering
\includegraphics[width=0.69\textwidth]{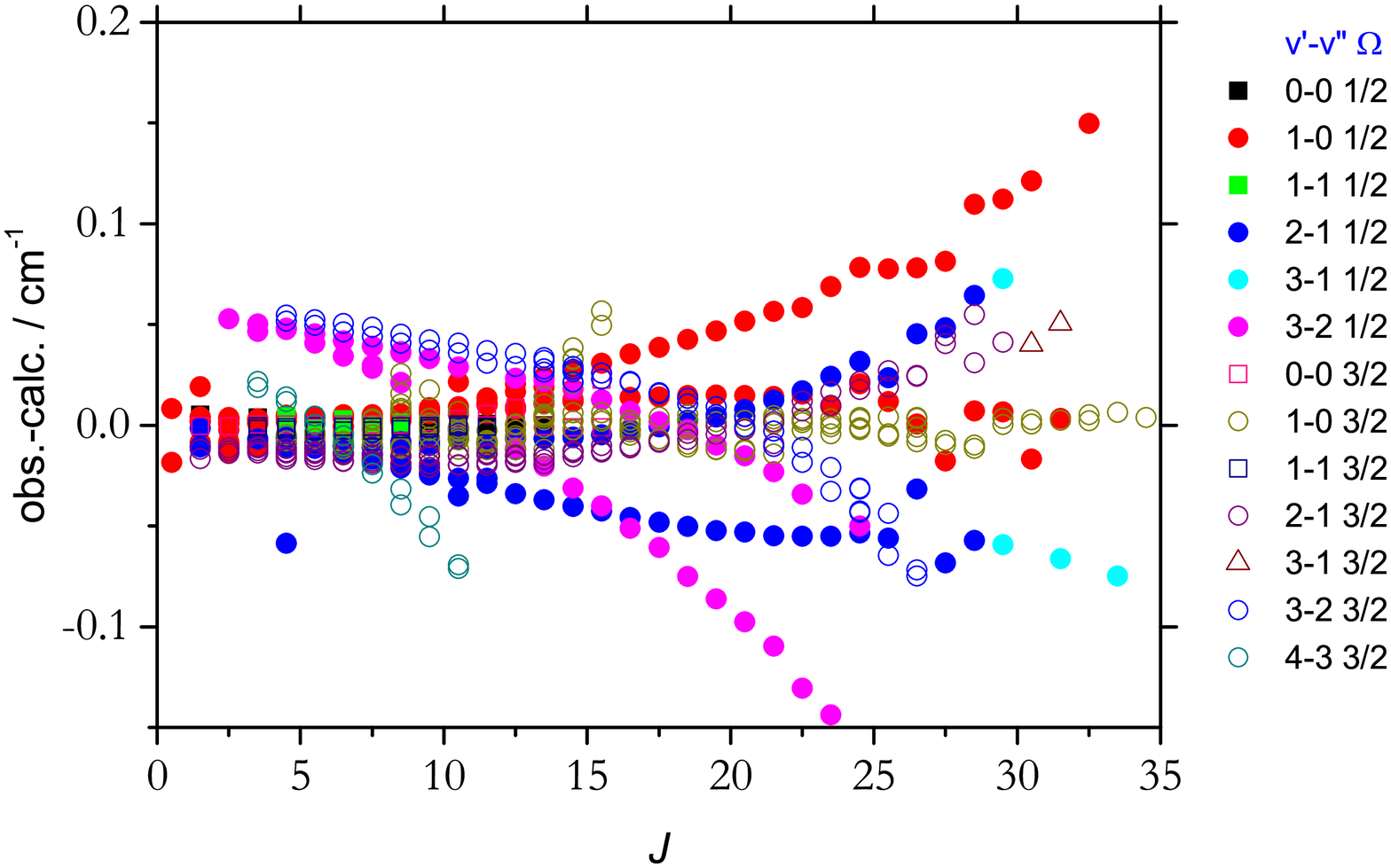}
\caption{Observed $-$ Calculated residuals for SH.}
\label{f:SH:obs-calc}
\end{figure}

The PEC, SO and EAM expansion parameters were obtained by fitting to
the experimental frequencies from the sources listed in
Table~\ref{t:HSlit} with a root-mean-square (rms) error of 0.03~\cm. The empirical vibrational information is limited to only $\Delta v=0$
and $\Delta v=1$ transitions with $v_{\rm max}=4$, which complicates obtaining a globally accurate model from the fitting. The refined curves are shown in
Figs~\ref{f:SH:PECs}--\ref{f:SH:Lx}. The quality of the fit is
illustrated in Fig.~\ref{f:SH:obs-calc}, where the Obs.--Calc.
residuals for all experimental data are shown and in Tables~\ref{t:obcalcSH-J} and \ref{t:obcalcSH-vib}. Most of the $\Delta v = 0 $ ($v_{\rm max} = 4$) and  $\Delta v = 1 $ ($v_{\rm max} = 1$) frequencies are reproduced within 0.005~\cm, except
for the $\Pi_{3/2}$ band, which is found to diverge at $J=25$ by about
0.15~\cm. Our final value for $D_{\rm e}$ (4.46~eV), corresponding to the best fit, is higher  than the  experimental  $X^{2}\Pi$ dissociation energy of
3.62$\pm 0.03$~eV ($D_{0}$) by \citet{91CoBaLe.SH},  as well as with
the \ai\ values recommended by \citet{03CsLeBu.SH} $D_{\rm e}$ = 3.791~eV and $D_{\rm 0}$ = 3.625~eV.
Therefore we limit our extrapolations to high vibrational excitations  to
those that do not to exceed this $D_{0}$ value.

\begin{table}
\caption{List of experimental data used in refinement of the SH \X\ potential energy curves.}
\begin{tabular}{lrcr}
\hline\hline
Source			& No. of transitions & Vibrational bands			& $J_{\rm max}$ \\ \hline
\citet{83BeAmWo.SH}	&	50	&	(1-0)				&	11.5	\\
\citet{84WiDaxx.SH}	&	285	&	(1-0), (2-1), (3-2)		&	34.5	\\
\citet{95RaBeEn.SH}	&	175	&	(1-0), (2-1), (3-2), (4-3)	&	16.5	\\
\citet{00YaKaRi.SH}	&	30	&	(1-0), (2-1), (3-2)		&	25.5	\\
\citet{11ElMaGu.SH}	&	6	&	(0-0)				&	4.5	\\
\citet{12MaElPi.SH} 	&	70	&	(0-0), (1-1)			&	16.5	\\ \hline
\end{tabular}
\label{t:HSlit}
\end{table}

\begin{table}
\caption{Experimental sources of NS spectroscopic data
 used in the refinement of the PEC. FTS=Fourier Transform Spectrometry, WS=millimetre and sub-millimetre Wave spectrometry. }
\begin{tabular}{llccc}
\hline\hline
 Study                  & Method        &       J       &       $\nu$   &       Range (\cm)    \\ \hline
\citet{86AnBoDa.NS}     & WS    &       2.5 -- 6.5       &       $(v,v)$, $v\leq\;5$       &       2.3 -- 10.1     \\
\citet{88SiBuHa.NS}     & FTS   &       0.5 -- 35.5      &       (1, 0)  &       1,149 -- 1,251     \\
\citet{95LeOzSa.NS}     & WS    &       0.5 -- 7.5       &       (0,0)      &      2.3 -- 11.6     \\
\hline
\end{tabular}
\label{t:NSlit}
 \end{table}


 The final spectroscopic model (provided as a \duo\ input file in the supplementary
 material) was then used to generate line lists for the following
 isotopologues: $^{32}$SH, $^{33}$H,$^{34}$H, $^{36}$SH ($J_{\rm max}
 = 64.5$) and $^{32}$SD ($J_{\rm max}=89.5$). In the \duo\
 calculations we used a sinc DVR method based on the grid of 501
 points equally distributed from 0.85 to 5.0 \AA. The equilibrium
 value of the \ai\ (MRCI/AWC5Z) dipole moment of SH (\X) is $\mu_e =
 0.801$~D (at $r = 1.3565$~\AA). The vibrationally averaged dipole
 moments, given by \( \mu_v = \bra{v}\mu\ket{v} \) where $\ket{v}$ is
 the vibrational eigenfunction of \X\ at the limit of $J=0$, are
 $\mu_0 = 0.794$~D and $\mu_1$ = -0.017~D.

The experimental value of the SH  ($v=0$) dipole moment  of  0.7580(1)~D was obtained by \citet{74MeDyxx.SH} in a Stark experiment. \citet{91BeFaGu.SH}
reported anomalously weak intensities of the fundamental band of SH
and obtained a very rough estimate for a relative dipole moment value
of $|\mu_1|/|\mu_0|$ of (0.011$\pm 0.016$~D)/0.63~D = 0.017$\pm 0.023$,
which compares favourably  to our value $|\mu_1|/|\mu_0| = 0.027$.






\begin{table}
\caption{Example of Observed $-$ Calculated residuals for SH frequencies as a function of $J$ for the (0, 0) band where $J'=J''+1$.} 
\begin{tabular}{cccrrr}
\hline
\hline
$J$ & $+/-$ & \multicolumn{1}{c}{$\Omega$} &  \multicolumn{1}{c}{Obs.} & \multicolumn{1}{c}{Calc.} & \multicolumn{1}{c}{Obs.-Calc.} \\
\hline
1.5	&	-	&	1.5	&	46.1289	&	46.1293	&	-0.0004	\\
2.5	&	+	&	1.5	&	64.5513	&	64.5519	&	-0.0006	\\
3.5	&	-	&	0.5	&	86.8367	&	86.8418	&	-0.0051	\\
3.5	&	+	&	1.5	&	82.9811	&	82.9808	&	0.0003	\\
4.5	&	-	&	0.5	&	106.2538	&	106.2508	&	0.0030	\\
5.5	&	-	&	1.5	&	119.6169	&	119.6179	&	-0.0010	\\
6.5	&	+	&	1.5	&	137.8762	&	137.8771	&	-0.0009	\\
7.5	&	+	&	1.5	&	156.1850	&	156.1846	&	0.0004	\\
8.5	&	-	&	0.5	&	181.9686	&	181.9684	&	0.0002	\\
9.5	&	+	&	0.5	&	200.5907	&	200.5909	&	-0.0002	\\
10.5	&	-	&	0.5	&	219.0671	&	219.0677	&	-0.0006	\\
11.5	&	+	&	1.5	&	228.2230	&	228.2231	&	-0.0001	\\
12.5	&	-	&	0.5	&	255.5412	&	255.5415	&	-0.0003	\\
13.5	&	-	&	0.5	&	273.5067	&	273.5068	&	-0.0001	\\
14.5	&	-	&	1.5	&	281.0630	&	281.0639	&	-0.0009	\\
15.5	&	+	&	1.5	&	298.3891	&	298.3901	&	-0.0010	\\
\hline\hline
\end{tabular}
\label{t:obcalcSH-J}
\end{table}

\begin{table}
\caption{Example of Observed $-$ Calculate residuals for SH frequencies for various vibrational bands as a function of $J$ where $J'=J''+1$.} 
\begin{tabular}{ccccrrr}
\hline \hline
$J$ & $+/-$  & \multicolumn{1}{c}{$\Omega$} & \multicolumn{1}{c}{Band} & \multicolumn{1}{c}{Obs.} & \multicolumn{1}{c}{Calc.} & \multicolumn{1}{c}{Obs.-Calc.} \\
\hline
1.5	&	+	&	0.5	&	(0, 0)	&	48.5370		&	48.5318	&	0.0052	\\
1.5	&	-	&	1.5	&	(1, 0)	&	2642.8296	&	2642.8262	&	0.0034	\\
1.5	&	+	&	0.5	&	(1, 0)	&	2644.8974	&	2644.8939	&	0.0035	\\
3.5	&	-	&	1.5	&	(1, 1)	&	80.5572		&	80.5590	&	-0.0018	\\
3.5	&	+	&	1.5	&	(1, 1)	&	80.5901		&	80.5907	&	-0.0006	\\
1.5	&	-	&	0.5	&	(2, 1)	&	2546.2628	&	2546.2735	&	-0.0107	\\
1.5	&	+	&	1.5	&	(2, 1)	&	2544.5747	&	2544.5867	&	-0.0120	\\
2.5	&	-	&	0.5	&	(3, 2)	&	2464.0678	&	2464.0150	&	0.0528	\\
3.5	&	-	&	0.5	&	(3, 2)	&	2479.3627	&	2479.3160	&	0.0467	\\
3.5	&	-	&	1.5	&	(4, 3)	&	2376.6704	&	2376.6516	&	0.0188	\\
3.5	&	+	&	1.5	&	(4, 3)	&	2376.6944	&	2376.6727	&	0.0217	\\
\hline\hline
\end{tabular}
\label{t:obcalcSH-vib}
\end{table}

\subsection{NS}

The electronic structure of the lowest seven states of NS was
intensely studied by \citet{13GaGaGo.NS}.  In this work we only concentrate on the
ground electronic state spectrum of NS.  The program MOLPRO
\citep{ MOLPRO} was used to compute \ai\ PEC, SOC and DMC for
the  NS \X\ ground state along with the spin-orbit coupling
curve for this state on a grid of 143 geometries distributed between
0.8 \AA\ and 2.7 \AA\ using the MRCI method and Douglas-Kroll Hamiltonian (dkroll=2)
with an aug-cc-pVQZ-DK basis set and the Davidson correction included. The  (2s,2p)/N and (2s,2p)/O complete active space (CAS)
is defined by 8330/4110 in the C$_{2v}$ symmetry employed by MOLPRO.

\begin{figure}
\centering
\includegraphics[width=0.49\textwidth]{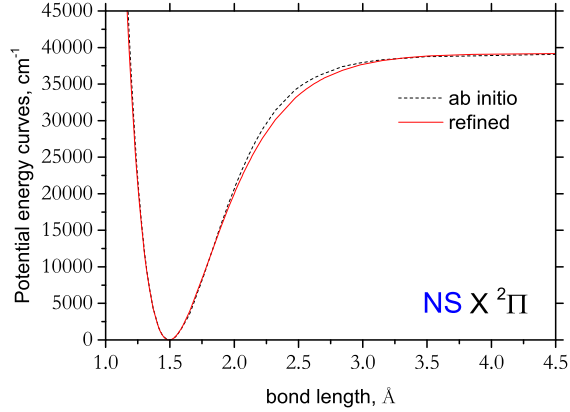}
\caption{\Ai\ PEC and refined PEC for the \X\ state of NS, see Eq.~(\ref{e:EMO}).}
\label{f:NS:pec}
\end{figure}

\begin{figure}
\centering
\includegraphics[width=0.49\textwidth]{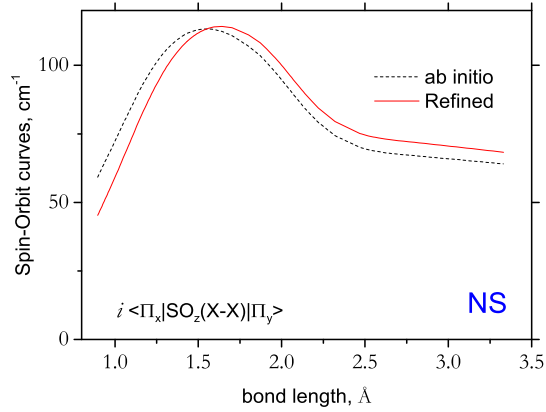}
\caption{\Ai\ and refined SO curves of NS.}
\label{f:NS:SO}
\end{figure}

\begin{figure}
\centering
\includegraphics[width=0.49\textwidth]{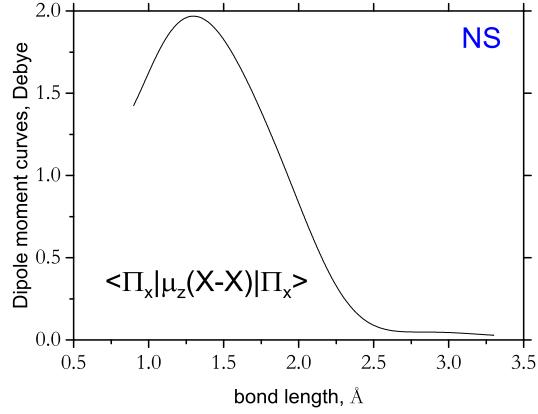}
\caption{\ai\ Dipole moment Curve of NS.}
\label{f:NS:dmc}
\end{figure}

The
\ai\ PEC, SO and DMC are shown in
Figs.~\ref{f:NS:pec}--\ref{f:NS:dmc}.  The \ai\ DMC of NS was modelled
using the damped-variable expansion in Eq.~\eqref{e:bob}.  The
equilibrium dipole value $\mu_{\rm e}$ is 1.834~D (at $r_{\rm e } =
1.494$~\AA), while the vibrationally averaged $\mu_0$ is 1.825~D, which is
in a good agreement with the experimental (Stark) value of 1.81~D
due to \citet{69AmSaHi.NS}.


In order to fit the PEC and SO curves of NS to the experimental data
for the \X\ state, several steps were taken. Firstly, the
program PGOPHER \citep{PGOPHER} was used to construct a list of
rovibronic energies using the molecular parameters published by
\citet{88SiBuHa.NS}. The MARVEL procedure \citep{MARVEL} was then used to
transform the list of measured experimental transitions summarised in
Table~\ref{t:NSlit} into several `networks' of derived energies:
these were used to check the rovibronic energies determined using
PGOPHER. This list of experimentally derived energies was then used to
perform an initial fit of the data, which was then improved by using
the actual experimentally measured frequencies.

The \duo\ calculations were based on the sinc DVR method comprising
701 points evenly distributed between 0.9 \AA\ and 3.3 \AA. The \ai\ \X\ PEC of
NS was represented using the EMO form in Eq.~\eqref{e:EMO} and refined
by fitting to 358 experimental frequencies covering the
rotational excitations up to $J=32.5$ and vibrational states up to $v=5$;
however, only $v=1-0$ transitions are for an the  vibration band, the remaining are microwave transitions for which $\Delta v=0$.

In the fits, we also included the 161 PGOPHER term values ($J\le 23.5$)
generated from the constants by
\citet{88SiBuHa.NS}.   Using experimentally-derived energies together with the
measured frequencies helps to constrain the fitted value to the absolute
energies, not only to the separation between them. This tends to make fits more
stable and prevent drifts between states (see also \citet{jt598}).
 In the refinement, the effects of the
spin-orbit coupling and $\Lambda$-doubling were taken into account:
the \ai\ SOC was morphed and the $\Lambda$-doubling curves was
refined by using the expression in Eq.~\eqref{e:bob}. The refined PEC
and SOC of NS are shown in Figs.~\ref{f:NS:pec} and \ref{f:NS:SO}.
The fitted parameters are presented in the supplementary material.

\begin{figure}
\centering
\includegraphics[width=0.69\textwidth]{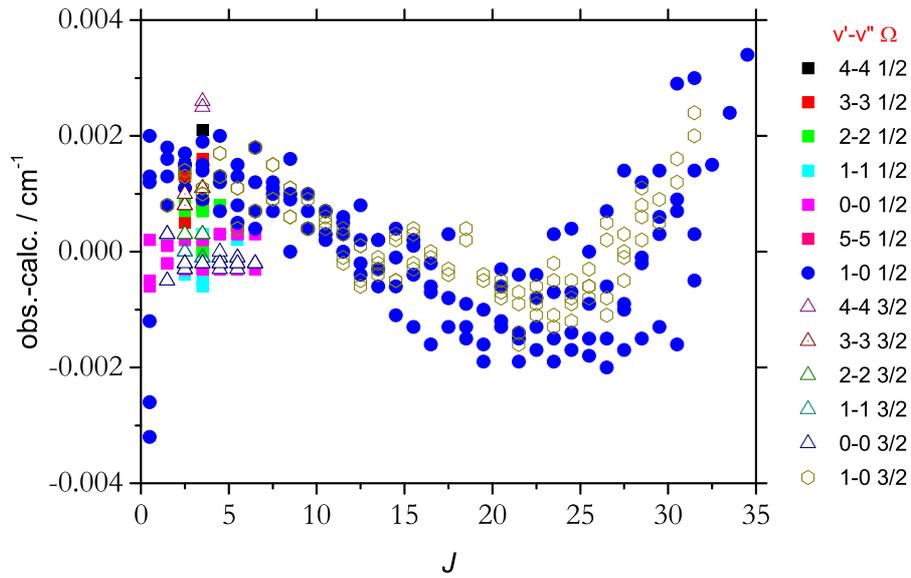}
\caption{Observed $-$ Calculated residuals for NS.}
\label{f:NS:obs-calc}
\end{figure}

Our final model reproduces the experimental frequencies with an rms error of 0.002~\cm. The experimentally derived energies are reproduced with an rms error of 0.03~\cm.
Fig.\ref{f:NS:obs-calc} shows the difference between the transition frequencies (\cm) calculated using the refined curves (Calc.) and those experimentally measured (Obs.). The error for most of the data is within 0.002~\cm, which is comparable to that obtained by the effective rotational methods.
Table~\ref{t:obcalcJ} shows a sample of the Obs.-Calc. residuals for the rotational energies ($v=0$) as a function of $J$, while Table~\ref{t:obcalcNU} compares some residuals for $v=0$ and $v=1$.

\begin{table}
\caption{Example of Observed minus calculated R-branch frequencies
as a function of $J$ for the NS (0, 0) band.}
\begin{tabular}{cccrrr}
\hline
\hline
$J$ & $+/-$ & \multicolumn{1}{c}{$\Omega$} & \multicolumn{1}{c}{Obs.} & \multicolumn{1}{c}{Calc.} & \multicolumn{1}{c}{Obs.-Calc.} \\
\hline
0.5	&	+	&	0.5	&	2.3026	&	2.3024	&	0.0002	\\
1.5	&	+	&	1.5	&	3.8760	&	3.8757	&	0.0003	\\
2.5	&	+	&	0.5	&	5.3808	&	5.3806	&	0.0002	\\
3.5	&	-	&	1.5	&	6.9758	&	6.9760	&	-0.0002	\\
3.5	&	-	&	0.5	&	6.9198	&	6.9196	&	0.0002	\\
4.5	&	+	&	1.5	&	8.5257	&	8.5259	&	-0.0002	\\
5.5	&	+	&	0.5	&	10.0095	&	10.0098	&	-0.0003	\\
6.5	&	+	&	0.5	&	11.5359	&	11.5356	&	0.0003	\\
\hline\hline
\end{tabular}
\label{t:obcalcJ}
\end{table}

\begin{table}
\caption{Example of Observed $-$ Calculated residuals for NS frequencies for various vibrational bands ($R$ branch).} 
\begin{tabular}{ccccrrr}
\hline
\hline
$J$ & $+/-$  & \multicolumn{1}{c}{$\Omega$} &  \multicolumn{1}{c}{Band} & \multicolumn{1}{c}{Obs.} & \multicolumn{1}{c}{Calc.} & \multicolumn{1}{c}{Obs.-Calc.} \\
\hline
0.5	&	+	&	0.5	&	(1, 0)	&	1206.5519	&	1206.5507	&	0.0012	\\
1.5	&	-	&	1.5	&	(1, 0)	&	1207.9205	&	1207.9196	&	0.0008	\\
2.5	&	+	&	0.5	&	(1, 0)	&	1209.5552	&	1209.5541	&	0.0011	\\
3.5	&	-	&	0.5	&	(1, 0)	&	1211.0378	&	1211.0369	&	0.0009	\\
4.5	&	+	&	0.5	&	(1, 0)	&	1212.5078	&	1212.5071	&	0.0007	\\
5.5	&	-	&	0.5	&	(1, 0)	&	1213.9651	&	1213.9647	&	0.0005	\\
6.5	&	+	&	0.5	&	(1, 0)	&	1215.4100	&	1215.4096	&	0.0004	\\
7.5	&	-	&	1.5	&	(1, 0)	&	1216.7633	&	1216.7624	&	0.0009	\\
8.5	&	+	&	0.5	&	(1, 0)	&	1218.2612	&	1218.2612	&	0.0000	\\
9.5	&	+	&	0.5	&	(1, 0)	&	1219.6812	&	1219.6802	&	0.0010	\\
10.5	&	+	&	1.5	&	(1, 0)	&	1221.0099	&	1221.0094	&	0.0005	\\
11.5	&	+	&	1.5	&	(1, 0)	&	1222.3989	&	1222.3991	&	-0.0002	\\
12.5	&	-	&	0.5	&	(1, 0)	&	1223.8239	&	1223.8231	&	0.0008	\\
13.5	&	+	&	1.5	&	(1, 0)	&	1225.1387	&	1225.1391	&	-0.0004	\\
14.5	&	-	&	1.5	&	(1, 0)	&	1226.4888	&	1226.4894	&	-0.0005	\\
15.5	&	+	&	1.5	&	(1, 0)	&	1227.8262	&	1227.8265	&	-0.0003	\\
17.5	&	-	&	0.5	&	(1, 0)	&	1230.4566	&	1230.4575	&	-0.0008	\\
18.5	&	+	&	0.5	&	(1, 0)	&	1231.7462	&	1231.7477	&	-0.0015	\\
20.5	&	+	&	1.5	&	(1, 0)	&	1234.3120	&	1234.3126	&	-0.0005	\\
22.5	&	-	&	0.5	&	(1, 0)	&	1236.7884	&	1236.7888	&	-0.0004	\\
24.5	&	+	&	1.5	&	(1, 0)	&	1239.2594	&	1239.2603	&	-0.0009	\\
25.5	&	+	&	1.5	&	(1, 0)	&	1240.4629	&	1240.4637	&	-0.0008	\\
28.5	&	-	&	0.5	&	(1, 0)	&	1243.9352	&	1243.9340	&	0.0012	\\
30.5	&	-	&	1.5	&	(1, 0)	&	1246.2757	&	1246.2745	&	0.0012	\\
32.5	&	+	&	0.5	&	(1, 0)	&	1248.4186	&	1248.4170	&	0.0015	\\
34.5	&	-	&	0.5	&	(1, 0)	&	1250.5983	&	1250.5935	&	0.0048	\\
2.5	&	-	&	0.5	&	(1, 1)	&	5.3493		&	5.3497	&	-0.0003	\\
3.5	&	-	&	0.5	&	(1, 1)	&	6.8637		&	6.8634	&	0.0003	\\
4.5	&	-	&	0.5	&	(1, 1)	&	8.4028		&	8.4025	&	0.0003	\\
5.5	&	+	&	0.5	&	(1, 1)	&	9.9289		&	9.9287	&	0.0002	\\
2.5	&	+	&	1.5	&	(2, 2)	&	5.3371		&	5.3368	&	0.0003	\\
3.5	&	+	&	0.5	&	(2, 2)	&	6.8201		&	6.8194	&	0.0007	\\
4.5	&	-	&	0.5	&	(2, 2)	&	8.3339		&	8.3331	&	0.0008	\\
2.5	&	-	&	0.5	&	(3, 3)	&	5.2622		&	5.2609	&	0.0013	\\
3.5	&	-	&	1.5	&	(3, 3)	&	6.8043		&	6.8032	&	0.0011	\\
2.5	&	+	&	0.5	&	(4, 4)	&	5.2046		&	5.2031	&	0.0015	\\
3.5	&	-	&	0.5	&	(4, 4)	&	6.6935		&	6.6914	&	0.0021	\\
\hline\hline
\end{tabular}
\label{t:obcalcNU}
\end{table}

\section{Results and Discussion}

\subsection{Line lists}

\subsubsection{SH}

\begin{table}
\caption{Statistics for the SH and NS line lists. }
\begin{tabular}{lrrrrr|rrrrr} \hline\hline
			& $^{32}$SH & $^{33}$SH & $^{34}$SH & $^{36}$SH & $^{32}$SD   & $^{14}$N$^{32}$S & $^{14}$N$^{33}$S & $^{14}$N$^{34}$S & $^{14}$N$^{36}$S  & $^{15}$N$^{32}$S	\\ \hline
$J_{\rm max}$		& 60.5	&	60.5	&	60.5	&	60.5	&	84.5	&	235.5	&	236.5	&	237.5	&	239.5	&	240.5	\\
number of energies	& 2326	&	2326	&	2328	&	2334	&	4532	&	31502	&	31802	&	32089	&	32620	&	33051	\\
number of lines		& 81,348	&	81,274	&	81,319	&	81,664	&	219,463	&	2,755,796	&	2,795,487&	2,831,482	&	2,901,113	&	2,957,016	\\
\hline\hline
\end{tabular}
\label{t:SHstats}
\end{table}

Line lists for the five most important isotopologues of SH were
computed using Duo. Table~\ref{t:SHstats} summarises the statistics
for these line lists. Those for SH contain almost 200,000 lines
while the heavier D atom means that the $^{32}$SD line list contains
more than double this number of transitions. In order to further improve the numerical stability of intensity calculations for high overtones,
we follow the procedure used by \cite{jt686} and apply a cutoff of $10^{-8}$~D to all matrix elements of the dipole moment $\bra{v}\mu\ket{v'}$.
The full line lists are
given in the ExoMol format \citep{jt631} as supplementary data.  Extracts
of the states and transitions files are given in
Tables~\ref{t:SHstates} and \ref{t:SHtrans} respectively. Apart from
the energy term values, statistical weights and quantum numbers, the
states file also contains the lifetimes \citep{jt624} and the Land\'{e} $g$-factors \citep{jt656}.

\begin{table*}
\centering
\caption{Extract from the states file of the $^{32}$S$^{1}$H line list.} 
\tt
\label{t:SHstates}
{\tt  \begin{tabular}{rrrrrrcclrrrr} \hline \hline
$n$ & Energy (\cm) & $g_i$ & $J$ & $\tau$ & $g$-factor	& Parity & e/f	& State	& $v$	&${\Lambda}$ &	${\Sigma}$ & $\Omega$ \\ \hline
1  & 360.537424		& 4 & 0.5 & inf		& -0.000697	& + &	e & X2Pi & 0	& 1 & -0.5	& 0.5	\\
2  & 2959.255407	& 4 & 0.5 & 0.708340	& -0.000699	& + &	e & X2Pi & 1	& 1 & -0.5	& 0.5	\\
3  & 5461.024835	& 4 & 0.5 & 0.247650	& -0.000699	& + &	e & X2Pi & 2	& 1 & -0.5	& 0.5	\\
4  & 7865.670867	& 4 & 0.5 & 0.130790	& -0.000700	& + &	e & X2Pi & 3	& 1 & -0.5	& 0.5	\\
5  & 10172.812283	& 4 & 0.5 & 0.083811	& -0.000699	& + &	e & X2Pi & 4	& 1 & -0.5	& 0.5	\\
6  & 12381.905056	& 4 & 0.5 & 0.060462	& -0.000699	& + &	e & X2Pi & 5	& 1 & -0.5	& 0.5	\\
7  & 14492.276445	& 4 & 0.5 & 0.047367	& -0.000697	& + &	e & X2Pi & 6	& 1 & -0.5	& 0.5	\\
8  & 16503.167217	& 4 & 0.5 & 0.039468	& -0.000694	& + &	e & X2Pi & 7	& 1 & -0.5	& 0.5	\\
9  & 18413.775027	& 4 & 0.5 & 0.034505	& -0.000687	& + &	e & X2Pi & 8	& 1 & -0.5	& 0.5	\\
10 & 20223.296263	& 4 & 0.5 & 0.031348	& -0.000679	& + &	e & X2Pi & 9	& 1 & -0.5	& 0.5	\\
11 & 21930.957245	& 4 & 0.5 & 0.029391	& -0.000669	& + &	e & X2Pi & 10	& 1 & -0.5	& 0.5	\\
12 & 23535.998338	& 4 & 0.5 & 0.028279	& -0.000659	& + &	e & X2Pi & 11	& 1 & -0.5	& 0.5	\\
13 & 25037.635167	& 4 & 0.5 & 0.027810	& -0.000646	& + &	e & X2Pi & 12	& 1 & -0.5	& 0.5	\\
14 & 26435.111738	& 4 & 0.5 & 0.027863	& -0.000627	& + &	e & X2Pi & 13	& 1 & -0.5	& 0.5	\\
15 & 27727.807976	& 4 & 0.5 & 0.028375	& -0.000599	& + &	e & X2Pi & 14	& 1 & -0.5	& 0.5	\\
\hline
\hline
\end{tabular}}
\mbox{}\\
{\flushleft
$n$:   State counting number.     \\
$\tilde{E}$: State energy in \cm. \\
$g_i$:  Total statistical weight, equal to ${g_{\rm ns}(2J + 1)}$.     \\
$J$: Total angular momentum.\\
$\tau$: Lifetime (s$^{-1}$).\\
$g$: Land\'{e} $g$-factors. \\
$+/-$:   Total parity. \\
$e/f$:   Rotationless parity. \\
State: Electronic state.\\
$v$:   State vibrational quantum number. \\
$\Lambda$:  Projection of the electronic angular momentum. \\
$\Sigma$:   Projection of the electronic spin. \\
$\Omega$:   Projection of the total angular momentum, $\Omega=\Lambda+\Sigma$. \\
}
\end{table*}

\begin{table}
\caption{Extract from the transitions file of the $^{32}$SH line list.}
\tt
\label{t:SHtrans}
\centering
\begin{tabular}{rrrr} \hline\hline
\multicolumn{1}{c}{$f$}	&	\multicolumn{1}{c}{$i$}	&
\multicolumn{1}{c}{$A_{fi}$ (s$^{-1}$)}	&\multicolumn{1}{c}{$\tilde{\nu}_{fi}$} \\ \hline
1037	&	1051	&	5.28E-006	&	12167.591629	\\
501	&	399	&	4.09E-006	&	12167.733027	\\
1625	&	1633	&	1.56E-005	&	12168.620635	\\
355	&	372	&	9.54E-006	&	12169.526762	\\
868	&	828	&	7.14E-003	&	12170.134555	\\
896	&	800	&	6.94E-003	&	12170.552236	\\
1064	&	1024	&	4.92E-006	&	12170.716947	\\
385	&	342	&	1.16E-005	&	12170.931908	\\
175	&	252	&	1.27E-005	&	12171.792693	\\
804	&	821	&	5.70E-007	&	12172.527689	\\
205	&	222	&	1.29E-005	&	12172.803180	\\
1528	&	1584	&	3.80E-004	&	12172.908871	\\
1510	&	1520	&	1.61E-005	&	12173.738888	\\
1552	&	1517	&	7.02E-006	&	12174.190774	\\
620	&	520	&	7.76E-006	&	12174.773993	\\
1551	&	1562	&	3.86E-004	&	12174.802908	\\
112	&	69	&	1.58E-008	&	12174.960464	\\
818	&	832	&	2.67E-005	&	12175.374485	\\
\hline\hline
\end{tabular}

\rm
\noindent
$f$: Upper  state counting number;\\
$i$:  Lower  state counting number; \\
$A_{fi}$:  Einstein-A coefficient in s$^{-1}$; \\
$\tilde{\nu}_{fi}$: transition wavenumber in \cm.\\
\end{table}
\subsubsection{NS}

For NS, five line lists were computed for the isotopologues
$^{14}$N$^{32}$S, $^{14}$N$^{33}$S, $^{14}$N$^{34}$S, $^{14}$N$^{36}$S
and $^{15}$N$^{32}$S (see Table~\ref{t:SHstats}).  The line lists are
based in the lowest $v_{\rm max} = $ 60 vibrational eigenfunctions
with the rotational quantum number $J$ ranging from 0.5 to 200.5 and the maximum energy term value $E_{\rm max}$ was
set from 0 \cm\ to 38~964.6 \cm. The
frequency window was set to 23~000~\cm\ which is just below the next electronic state, $a$~$^{4}\Pi$ \citep{13GaGaGo.NS}. The values of $v_{\rm
  max}$ and $E_{\rm max}$ correspond to the dissociation limit of
4.83~eV determined by \citep{04CzViXX.NS}.  
A dipole moment cutoff of $10^{-8}$~D was also used.
Again, Tables~\ref{t:SNstates} and \ref{t:SNtrans} show extracts from
the corresponding states and transition files.

\begin{table}
\centering
\caption{Extract from the states file for the line list of $^{14}$N$^{32}$S. } 
\tt
\label{t:SNstates}
\begin{tabular}{crcccccrrrrrcr}
\hline\hline
$i$ & Energy (\cm) & $g_i$ & $J$ & $\tau$ & $g$ & Parity  & e/f & State   & $v$ & ${\Lambda}$ & ${\Sigma}$ &${\Omega}$   \\
\hline
1  & 0.000000		& 6 & 0.5 & inf		& -0.000767 &	+ & e &	X2Pi	& 0	& 1	& -0.5	& 0.5	\\
2  & 1204.267014	& 6 & 0.5 & 7.5810E-001	& -0.000767 &	+ & e &	X2Pi	& 1	& 1	& -0.5	& 0.5	\\
3  & 2391.811399	& 6 & 0.5 & 3.7574E-001	& -0.000767 &	+ & e &	X2Pi	& 2	& 1	& -0.5	& 0.5	\\
4  & 3562.543092	& 6 & 0.5 & 2.4892E-001	& -0.000767 &	+ & e &	X2Pi	& 3	& 1	& -0.5	& 0.5	\\
5  & 4716.365128	& 6 & 0.5 & 1.8592E-001	& -0.000767 &	+ & e &	X2Pi	& 4	& 1	& -0.5	& 0.5	\\
6  & 5853.187808	& 6 & 0.5 & 1.4840E-001	& -0.000767 &	+ & e &	X2Pi	& 5	& 1	& -0.5	& 0.5	\\
7  & 6972.935476	& 6 & 0.5 & 1.2360E-001	& -0.000767 &	+ & e &	X2Pi	& 6	& 1	& -0.5	& 0.5	\\
8  & 8075.513299	& 6 & 0.5 & 1.0604E-001	& -0.000767 &	+ & e &	X2Pi	& 7	& 1	& -0.5	& 0.5	\\
9  & 9160.809357	& 6 & 0.5 & 9.3014E-002	& -0.000767 &	+ & e &	X2Pi	& 8	& 1	& -0.5	& 0.5	\\
10 & 10228.711651	& 6 & 0.5 & 8.2993E-002	& -0.000767 &	+ & e &	X2Pi	& 9	& 1	& -0.5	& 0.5	\\
11 & 11279.122350	& 6 & 0.5 & 7.5073E-002	& -0.000767 &	+ & e &	X2Pi	& 10	& 1	& -0.5	& 0.5	\\
12 & 12311.955993	& 6 & 0.5 & 6.8677E-002	& -0.000767 &	+ & e &	X2Pi	& 11	& 1	& -0.5	& 0.5	\\
13 & 13327.125579	& 6 & 0.5 & 6.3426E-002	& -0.000767 &	+ & e &	X2Pi	& 12	& 1	& -0.5	& 0.5	\\
14 & 14324.536677	& 6 & 0.5 & 5.9056E-002	& -0.000767 &	+ & e &	X2Pi	& 13	& 1	& -0.5	& 0.5	\\
15 & 15304.087622	& 6 & 0.5 & 5.5383E-002	& -0.000767 &	+ & e &	X2Pi	& 14	& 1	& -0.5	& 0.5	\\
16 & 16265.666432	& 6 & 0.5 & 5.2272E-002	& -0.000767 &	+ & e &	X2Pi	& 15	& 1	& -0.5	& 0.5	\\
17 & 17209.152585	& 6 & 0.5 & 4.9619E-002	& -0.000767 &	+ & e &	X2Pi	& 16	& 1	& -0.5	& 0.5	\\
18 & 18134.433115	& 6 & 0.5 & 4.7350E-002	& -0.000767 &	+ & e &	X2Pi	& 17	& 1	& -0.5	& 0.5	\\
19 & 19041.403694	& 6 & 0.5 & 4.5404E-002	& -0.000767 &	+ & e &	X2Pi	& 18	& 1	& -0.5	& 0.5	\\
20 & 19929.953675	& 6 & 0.5 & 4.3734E-002	& -0.000767 &	+ & e &	X2Pi	& 19	& 1	& -0.5	& 0.5	\\
\hline\hline
\end{tabular}
\mbox{}\\
{\flushleft
$i$:   State counting number.     \\
$\tilde{E}$: State energy in \cm. \\
$g$:  Total statistical weight, equal to ${g_{\rm ns}(2J + 1)}$.     \\
$J$: Total angular momentum.\\
$\tau$: Lifetime (s$^{-1}$).\\
$g$: Land\'{e} $g$-factors. \\
$+/-$:   Total parity. \\
$e/f$:   Rotationless parity. \\
State: Electronic state.\\
$v$:   State vibrational quantum number. \\
$\Lambda$:  Projection of the electronic angular momentum. \\
$\Sigma$:   Projection of the electronic spin. \\
$\Omega$:   $\Omega=\Lambda+\Sigma$, projection of the total angular momentum.\\
}
\end{table}

\begin{table}
\centering
\caption{Extract from the transition file for the line list of $^{14}$N$^{32}$S.} 
\tt
\label{t:SNtrans}
\begin{tabular}{rrrr}
\hline
\multicolumn{1}{c}{$f$} &       \multicolumn{1}{c}{$i$} & \multicolumn{1}{c}{$A_{fi}$ (s$^{-1}$)}       &\multicolumn{1}{c}{$\tilde{\nu}_{fi}$} \\
\hline\hline
7738	&	7381	&	1.0816E-12	&	21129.956080	\\
5205	&	5270	&	1.5211E-12	&	21129.964779	\\
5098	&	5376	&	1.5212E-12	&	21129.965742	\\
11335	&	11591	&	3.1996E-13	&	21129.967612	\\
13834	&	13865	&	1.5996E-13	&	21129.972884	\\
6281	&	5916	&	1.5599E-14	&	21129.973205	\\
7633	&	7486	&	1.0819E-12	&	21129.974657	\\
17595	&	17799	&	1.3308E-13	&	21130.010597	\\
18641	&	18492	&	2.9298E-12	&	21130.032427	\\
14941	&	14800	&	1.0700E-15	&	21130.038476	\\
15071	&	14910	&	3.7284E-14	&	21130.042388	\\
22202	&	21929	&	5.7979E-15	&	21130.049792	\\
7889	&	7931	&	7.6269E-17	&	21130.090860	\\
12655	&	12499	&	4.4455E-18	&	21130.105215	\\
5327	&	5386	&	7.8126E-18	&	21130.120789	\\
21381	&	21404	&	7.3821E-16	&	21130.145157	\\
13229	&	12887	&	5.8001E-15	&	21130.148827	\\
25770	&	25777	&	3.0568E-15	&	21130.159305	\\
18726	&	18407	&	2.9242E-12	&	21130.204387	\\
12449	&	12495	&	2.5432E-13	&	21130.208111	\\
15334	&	15370	&	2.3191E-13	&	21130.223604	\\
\hline\hline
\end{tabular}

\mbox{}\\
\noindent
 $f$: Upper  state counting number;\\
$i$:  Lower  state counting number; \\
$A_{fi}$:  Einstein-A coefficient in s$^{-1}$; \\
$\tilde{\nu}_{fi}$: transition wavenumber in \cm.\\
\end{table}

\subsection{Partition Functions}

Partition functions were computed up to 5000~K for every species considered at 1 K intervals. These can be found in the supplementary material. We have also fitted the partition function to the function form of \citet{jt263}:
\begin{equation}\label{eq:fit1d}
\log_{10}Q(T) = \sum\limits^{9}_{n=0} a_n (\log_{10}T)^n.
\end{equation}
Table~\ref{table:pf} gives the expansion coefficients for the parent isotopologues, fits for
other species can be found in the supplementary material, which reproduce the ExoMol partition function within 2--3~\%.

In general our partition functions are in excellent agreement with
those available from other sources, namely from
\citet{84SaTaxx.partfunc}, \citet{16BaCoxx.partfunc} and the CDMS
database \citep{CDMS}, once allowance is made for the nuclear spin
conventions employed: ExoMol uses the HITRAN convention which leads to
a factor of 2 for $^{32}$SH and 3 for $^{14}$N$^{32}$S compared to the
`astronomers' convention employed by \citet{84SaTaxx.partfunc} and
\citet{16BaCoxx.partfunc}. The only significant disagreement is for SH below
1000~K, where the results of \citet{84SaTaxx.partfunc} follow the wrong
trend; \citet{84SaTaxx.partfunc} only aimed to be accurate above
1000~K.

Our partition functions should be complete up to 5000~K. For SH,  the completeness is within 0.3~\%, which corresponds to the number of states in our line list above the experimental dissociation energy \citep{91CoBaLe.SH} at $T=5000$~K. For NS, we mainly miss the contributions from the $a^{4}$~$\Pi$ rovibronic states not considered here ($T_{\rm e} = $ 24,524~\cm, \citet{13GaGaGo.NS}). This should  not exceed 1~\%\ judging by the contribution to $Q(T)$ from the \X\ energies of NS above 24,524~\cm.

\begin{table} 
\centering
\caption{Expansion coefficients for the partition function given
by Eq.~(\ref{eq:fit1d}). Parameters for other isotopologues can be found in the
supplementary material.}
 \begin{tabular}{crr}
\hline\hline
$a_i$	&	$^{32}$SH	&	$^{14}$N$^{32}$S \\ \hline
$a_0$	&	1.20403		&	1.10562		\\
$a_1$	&	-0.47064	&	0.19180		\\
$a_2$	&	2.88449		&	1.01142		\\
$a_3$	&	-6.62867	&	-1.40930	\\
$a_4$	&	7.73660		&	1.96534		\\
$a_5$	&	-5.21848	&	-1.82956	\\
$a_6$	&	2.16261		&	1.00323		\\
$a_7$	&	-0.54158	&	-0.31160	\\
$a_8$	&	0.07488		&	0.05081		\\
$a_9$	&	-0.00437	&	-0.00338	\\
\hline \hline
\end{tabular}
\label{table:pf}
\end{table}


\begin{figure}
\centering
\includegraphics[width=220pt]{SH_pf_v4.eps}
\includegraphics[width=220pt]{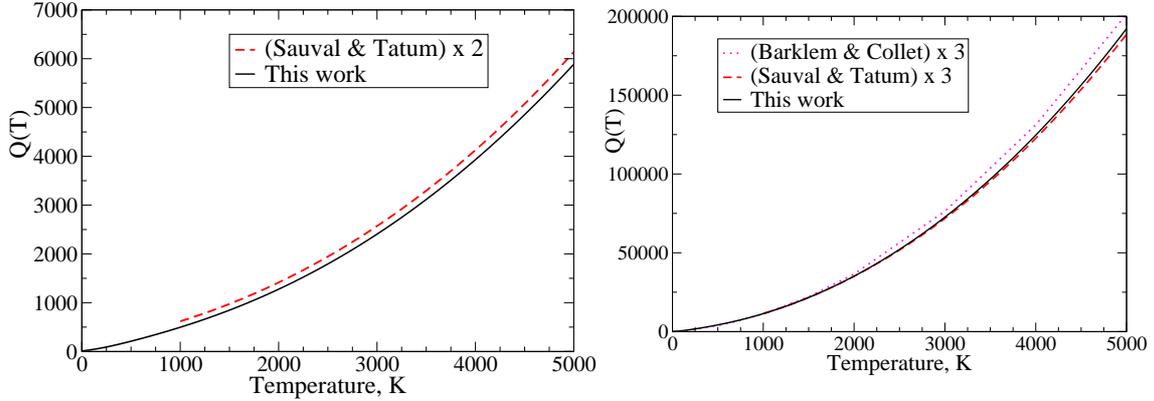}
\caption{Temperature dependence of the partition functions of SH (left) and NS (right) computed using our line lists and compared to those by \protect\citet{84SaTaxx.partfunc} and \protect\citet{16BaCoxx.partfunc}.}
\label{f:pf}
\end{figure}

\subsection{Spectra}

\subsubsection{SH}

\begin{figure}
\centering
\includegraphics[width=320pt]{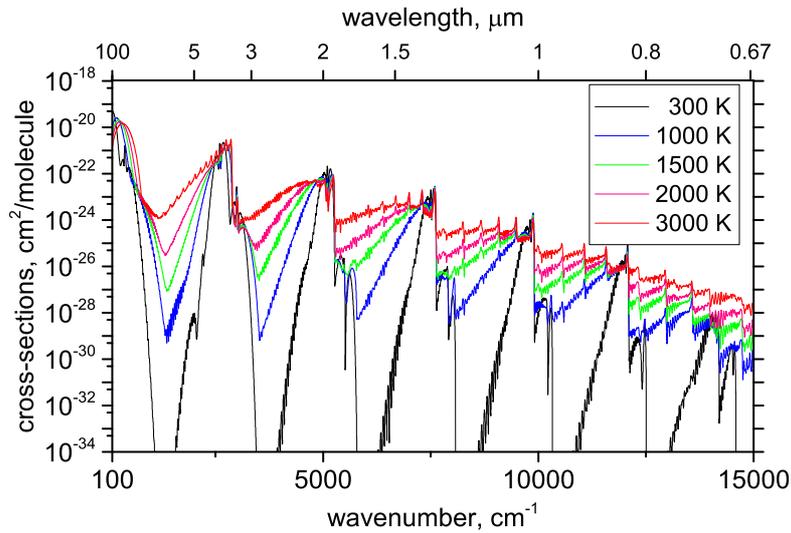}
\caption{Temperature dependence of our simulated SH absorption spectrum. A Gaussian line profile with HWHM=1~\cm\ is used. The curves become flatter with increasing temperature.}
\label{f:SH:temp}
\end{figure}

\begin{figure}
\centering
\includegraphics[width=320pt]{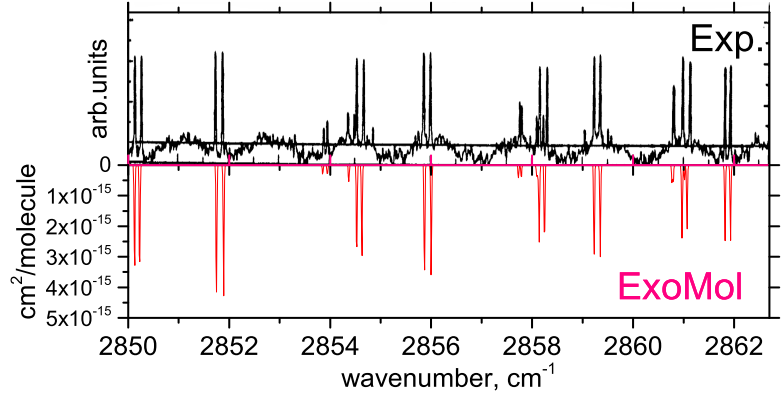}
\caption{Comparison of simulated spectra at 2000 K (HWHM=0.01~\cm) using the  new ExoMol line list for $^{32}$SH and experimental spectra by \citet{84WiDaxx.SH}.}
\label{f:2000:SH:emiss}
\end{figure}

\begin{figure}
\centering
\includegraphics[width=220pt]{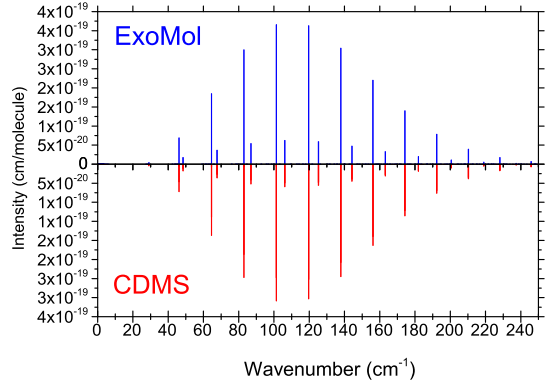}
\caption{Comparison of simulated spectra using the new ExoMol line list for $^{32}$SH and the CDMS database for the $\nu=0$ band.}
\label{f:SH:CDMS}
\end{figure}

Figure~\ref{f:SH:temp} gives an overview of the spectrum of SH at
different temperatures. Note how intensities drop exponentially across the entire frequency range shown:
no unphysical, plateau-like structures at higher frequencies is present \citep{16MeMeSt}.
This illustrates that our measures to prevent this spurious effect were successful. Figure~\ref{f:2000:SH:emiss} compares a
simulated emission spectrum at $T=2000$~K (HWHW = 0.01~\cm) to the
experimental spectrum of \citet{84WiDaxx.SH}.  The agreement is good,
especially considering the complex coupling and the limited amount of
the experimental data.  Figure~\ref{f:SH:CDMS} shows a comparison of
an SH spectrum at $T=296$~K computed using the ExoMol line list with
that from the CDMS database~\citep{CDMS}.  The CDMS spectrum was
obtained using the equilibrium dipole moment of 0.7580~D from
\citet{74MeDyxx.SH}, while our value is 0.794~D.


\subsubsection{NS}

\begin{figure}
\centering
    \includegraphics[width=0.49\textwidth]{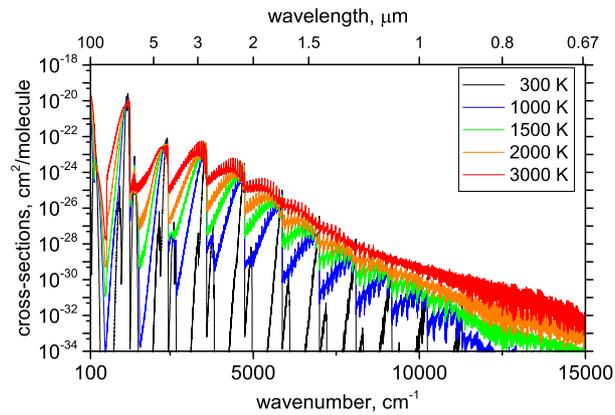}
\caption{Temperature  dependence of our simulated NS absorption spectrum. A Gaussian line profile with HWHM=1~\cm\ is used. The curves become flatter with increasing temperature.}
\label{f:NStemp}
\end{figure}

\begin{figure}
\centering
\includegraphics[width=0.45\textwidth]{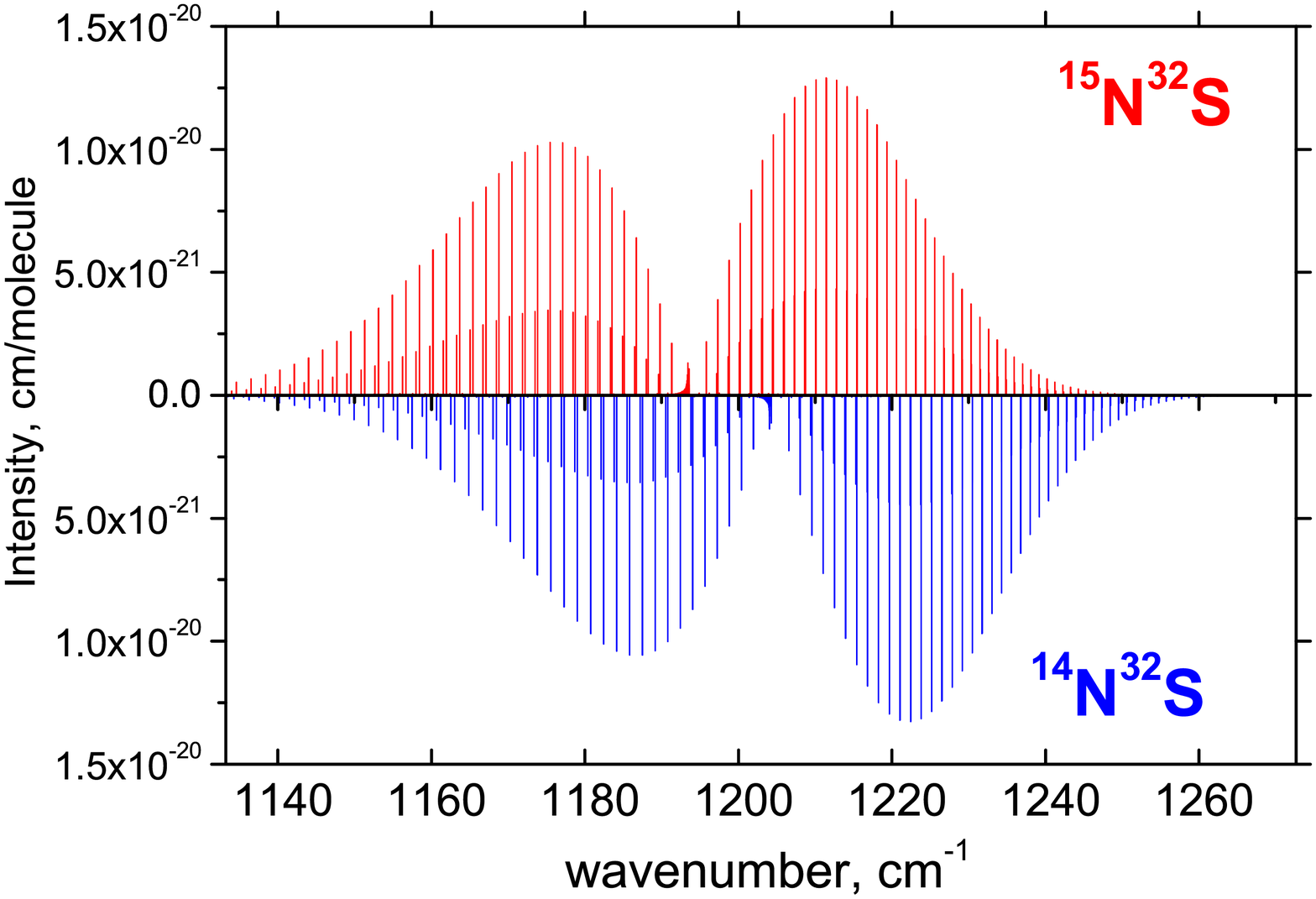}
\includegraphics[width=0.45\textwidth]{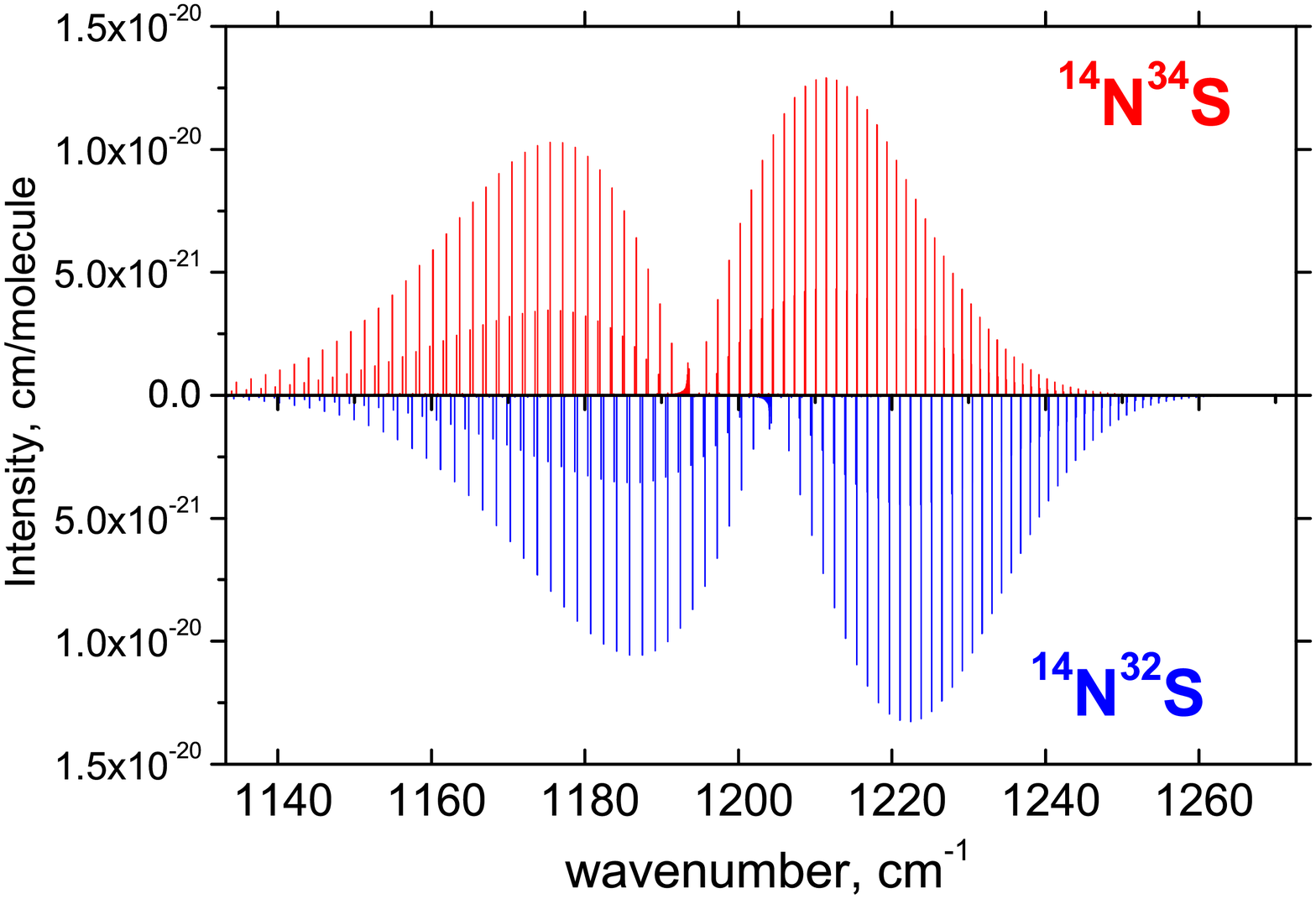}
\caption{Comparison of spectra (298~K) for the fundamental
$v =1- 0$  band of NS for $^{15}$N$^{32}$S (left)  and  $^{14}$N$^{34}$S (right)
against $^{14}$N$^{32}$S.}
\label{f:NSiso1}
\end{figure}

 Figure~\ref{f:NStemp} shows a comparison of spectra for the main isotopologue of NS as a function of temperature. Figure
\ref{f:NSiso1}   shows the effects on the spectra of NS when the
main isotopes of N and S are substituted. It can be seen that
effect of substitution of the N atom leads to redshifts of up to 28~\cm\ and that of substituting S is up to 10~\cm.

\begin{figure}[H]
\centering
\includegraphics[width=0.45\textwidth]{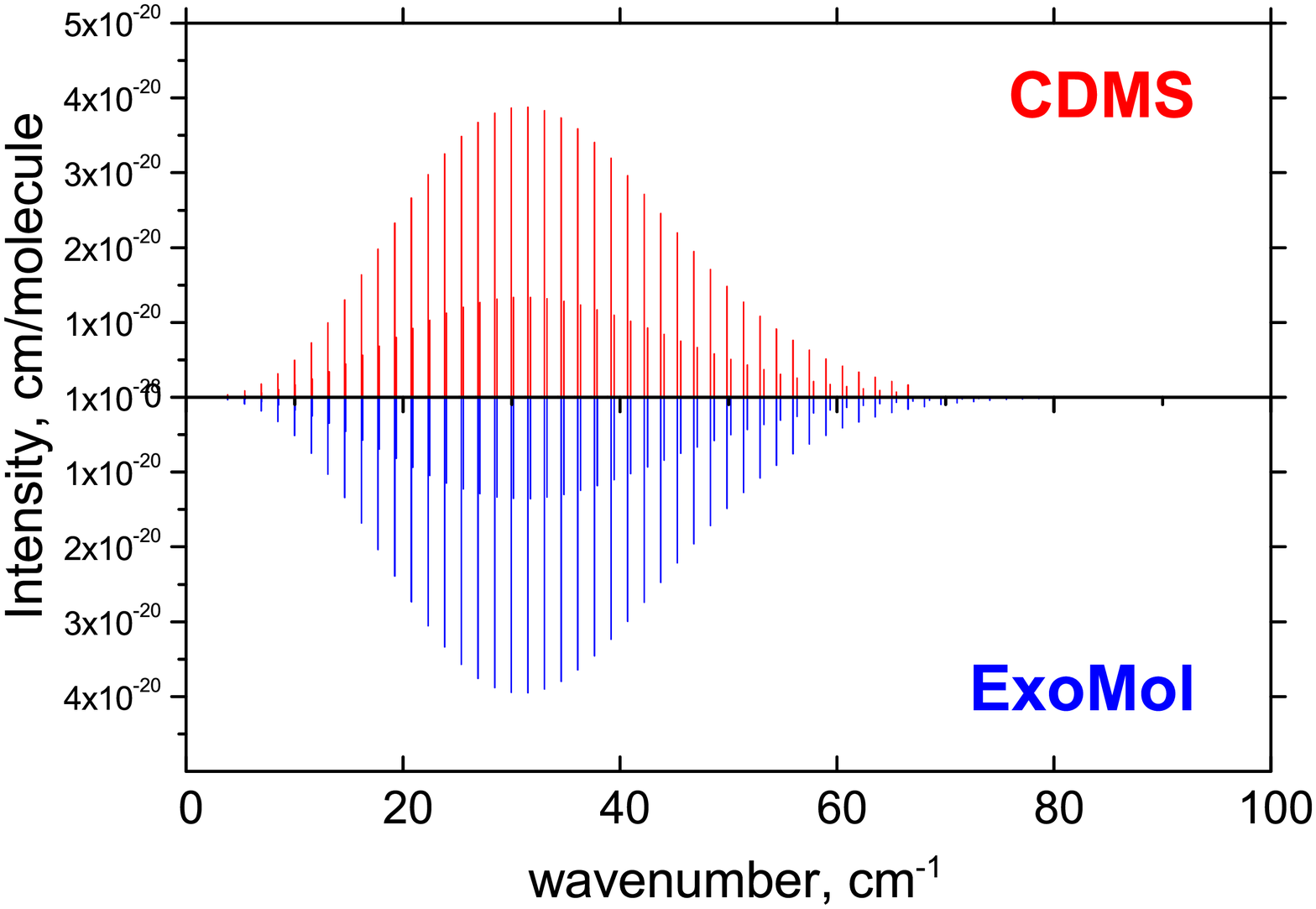}
\includegraphics[width=0.45\textwidth]{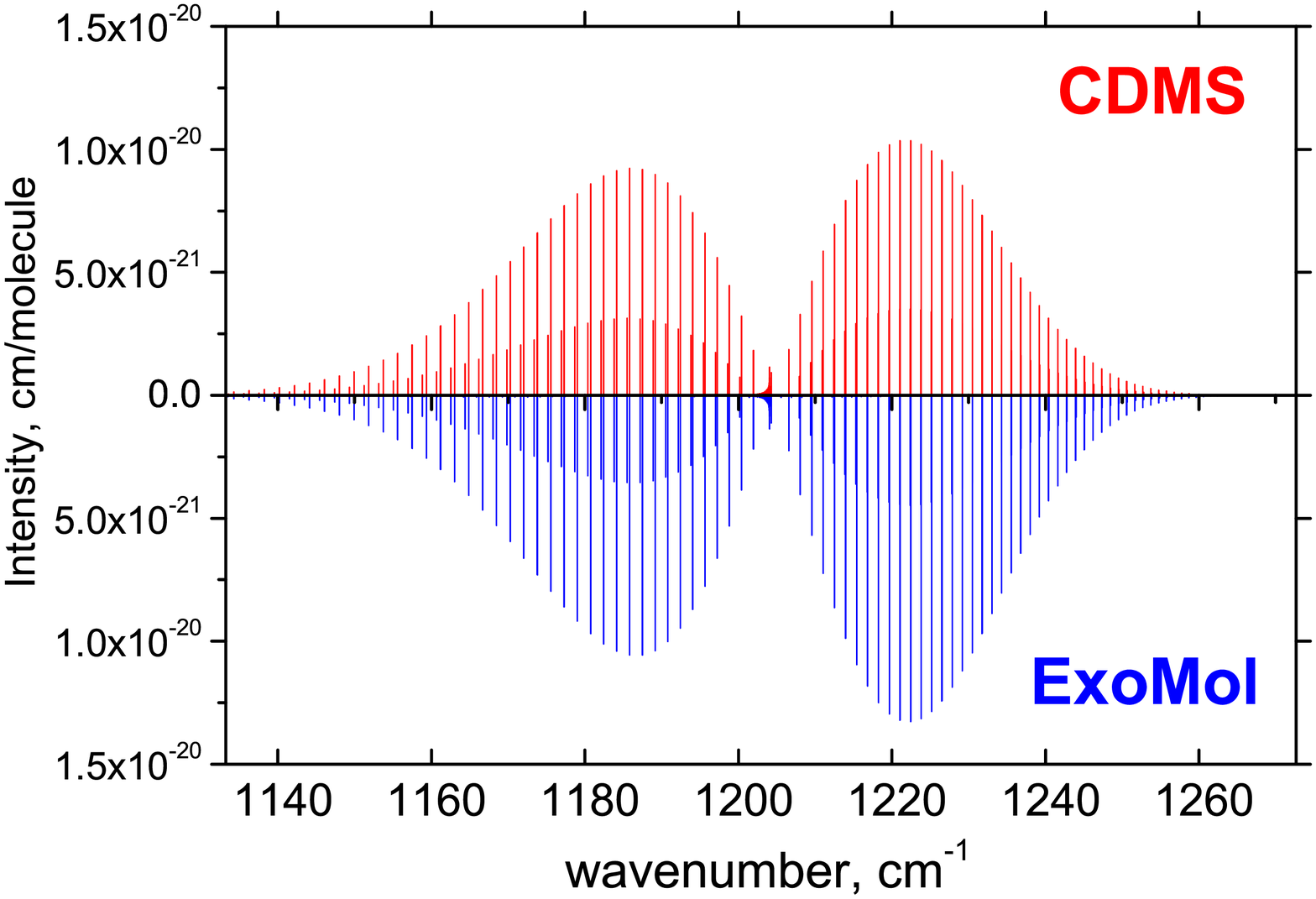}
\caption{Detailed Comparison our results with those given by CDMS \citep{CDMS} for pure rotational
transitions (left) and the $v =1- 0$ fundamental band (right).}
\label{f:SNcdms-2}
\end{figure}

To illustrate the accuracy of our line lists,
spectra have been simulated and compared to the existing CDMS database \citep{CDMS} for the rotational and fundamental bands of NS, see Fig.\ref{f:SNcdms-2}. In order to make this comparison, hyperfine splitting was averaged in the CDMS transitions of NS. Our intensities are in good agreement with those from CDMS.  Some difference between the intensities from the fundamental band is due to the different \ai\ dipole moments used: our transition dipole $\mu_1$ for the fundamental band is 0.049~D, while CDMS used $\mu_1 = 0.045$~D from unpublished work by H.~M\"{u}ller.

\subsection{Lifetimes}

\begin{figure}
\centering
\includegraphics[width=0.5\textwidth]{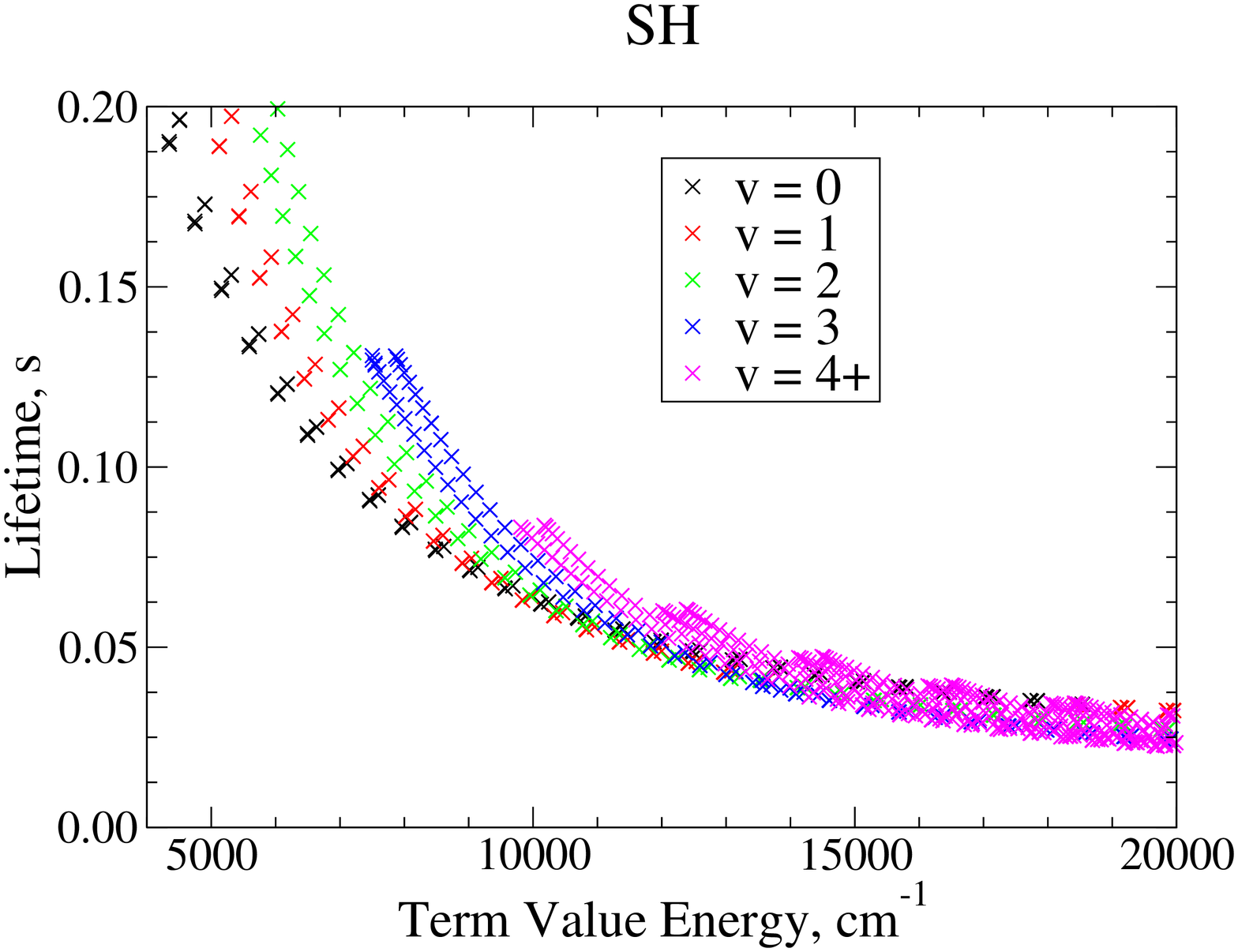} 
\includegraphics[width=0.49\textwidth]{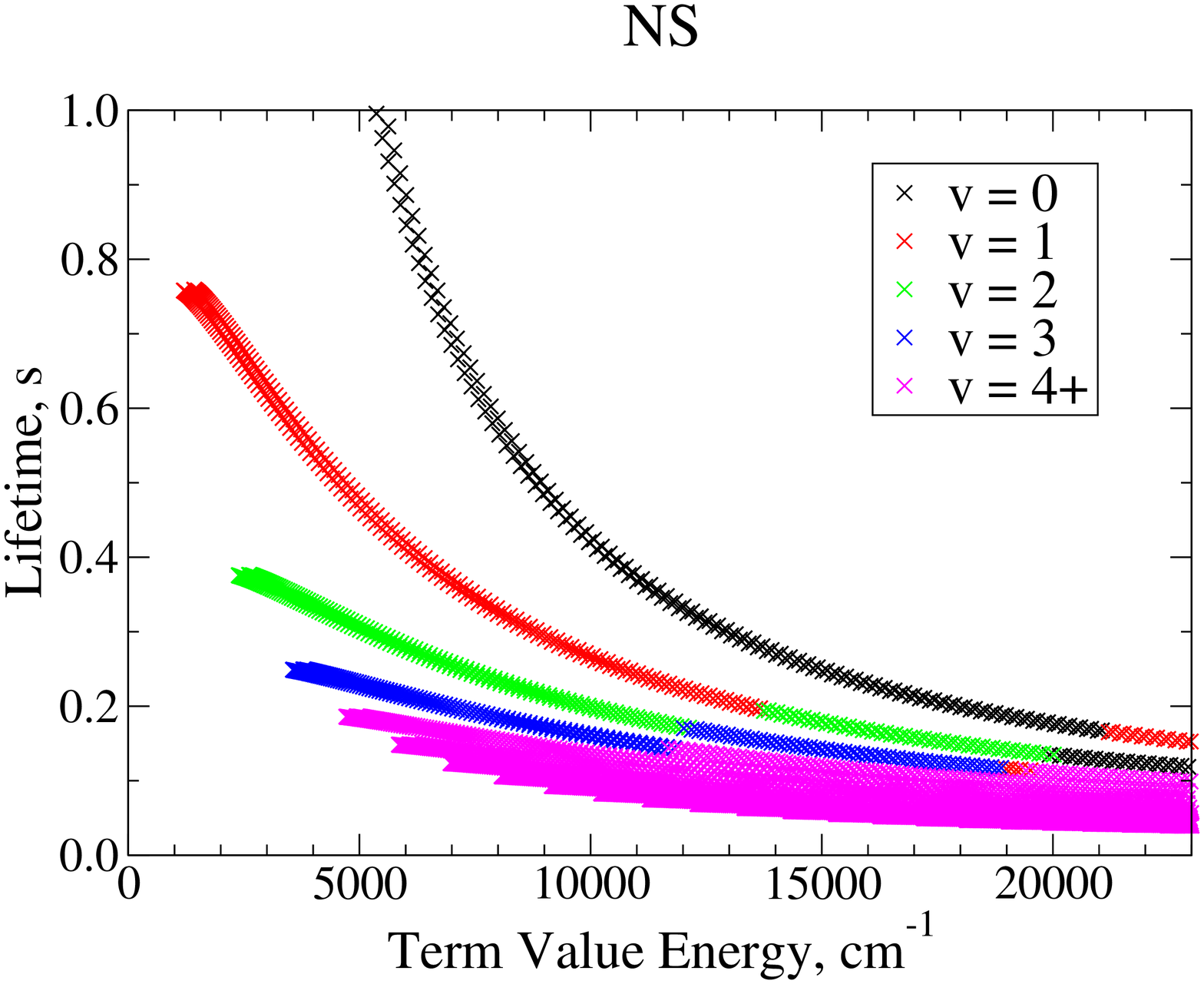} 
\caption{Lifetimes calculated for levels of  $^{32}$SH and $^{14}$N$^{32}$S  in their the \X\ electronic states. For SH the lifetimes increase with
vibrational excitation; for NS the longest lived states are for $v=0$ and
the lifetimes decrease with vibrational excitation.}
\label{f:lifetime}
\end{figure}

Lifetimes for states within the \X\ ground state can be computed in straightforward manner from our line lists \citep{jt624}. These are included in states file, see Tables~\ref{t:SHstates} and \ref{t:SNstates} above. Figure \ref{f:lifetime} presents lifetimes states associated with the main isotopologues of SH and NS. As can be seen for NS, lifetimes for each vibrational state decrease by approximately up to one order of magnitude as energy is increased.

\section{Conclusions}

New line lists for the electronic ground states of major isotopologues
of both SH and NS are generated using a high level \ai\ theory
refined to available experimental data. For SH, each line list contains approximately 81~000
transitions and 2300 states, with a range up to the dissociation limit
of 29~234 cm$^{-1}$, vibrational coverage up to 14 and rotational
coverage to $J=60.5$.   For NS, the line lists
contain 2.7 -- 2.9 million transitions up to 23~000~\cm\ and 31~000 -- 33~000 states with a
range up to the dissociation limit of 38~964.6 cm$^{-1}$, with
vibrational coverage up to $\nu$=54 and rotational coverage to
$J=235.5$.  These are the only available hot line lists for
these molecules.  The line lists, which are named SNaSH, are available
from the CDS (http://cdsarc.u-strasbg.fr) and ExoMol (www.exomol.com)
data bases.

Our model is affected by the limitations of the experimental data as well as by the \ai\ accuracy, especially for the dipole moment calculations. The latter is critical for accurate retrievals of molecular opacities in different astronomical bodies. It is typical for hot diatomics that experimental data on the dipole moments are either absent or extremely inaccurate, which emphasises the importance of the \ai\ calculations of this property.

Considering the importance of the SH molecule for star spectroscopy, it would be important to extend the present study with the inclusion of the \A\ electronic state, which would allow accurate modelling and prediction in the high-energy visible and UV spectral regions.

Our new line lists should enable detection of and inclusion in models
of SH and NS for exoplanet temperatures which are largely not covered
by experimental results. The ExoMol project has already provided line lists for
for several sulphur-containing molecules, namely
CS \citep{jt615}, SiS \citep{jt724}, PS \citep{jt703}, H$_2$S \citep{jt640}, SO$_2$ \citep{jt635} and SO$_3$ \citep{jt641}. SO is probably the only
major sulphur-baring species which is important at elevated temperatures
which is missing from this list.

\section{Acknowledgements}

This work was supported by the UK Science and Technology Research Council (STFC)
No. ST/M001334/1 and the COST action MOLIM No. CM1405.  This work made extensive
use of UCL's Legion  high performance computing facility.

\bibliographystyle{mnras}

\begin{thebibliography}{}
\makeatletter
\relax
\def\mn@urlcharsother{\let\do\@makeother \do\$\do\&\do\#\do\^\do\_\do\%\do\~}
\def\mn@doi{\begingroup\mn@urlcharsother \@ifnextchar [ {\mn@doi@}
  {\mn@doi@[]}}
\def\mn@doi@[#1]#2{\def\@tempa{#1}\ifx\@tempa\@empty \href
  {http://dx.doi.org/#2} {doi:#2}\else \href {http://dx.doi.org/#2} {#1}\fi
  \endgroup}
\def\mn@eprint#1#2{\mn@eprint@#1:#2::\@nil}
\def\mn@eprint@arXiv#1{\href {http://arxiv.org/abs/#1} {{\tt arXiv:#1}}}
\def\mn@eprint@dblp#1{\href {http://dblp.uni-trier.de/rec/bibtex/#1.xml}
  {dblp:#1}}
\def\mn@eprint@#1:#2:#3:#4\@nil{\def\@tempa {#1}\def\@tempb {#2}\def\@tempc
  {#3}\ifx \@tempc \@empty \let \@tempc \@tempb \let \@tempb \@tempa \fi \ifx
  \@tempb \@empty \def\@tempb {arXiv}\fi \@ifundefined
  {mn@eprint@\@tempb}{\@tempb:\@tempc}{\expandafter \expandafter \csname
  mn@eprint@\@tempb\endcsname \expandafter{\@tempc}}}

\bibitem[\protect\citeauthoryear{Amano, Saito, Hirota  \& Morino}{Amano
  et~al.}{1969}]{69AmSaHi.NS}
Amano T.,  Saito S.,  Hirota E.,   Morino Y.,  1969, \mn@doi [J. Mol.
  Spectrosc.] {10.1016/0022-2852(69)90145-3}, 32, 97

\bibitem[\protect\citeauthoryear{Anacona}{Anacona}{1994}]{94Anacon.NS}
Anacona J.~R.,  {1994}, \mn@doi [Spectra Chimica Acta A]
  {{10.1016/0584-8539(94)80139-8}}, {50}, 909

\bibitem[\protect\citeauthoryear{Anacona}{Anacona}{1995}]{95Anacon.NS}
Anacona J.~R.,  {1995}, \mn@doi [Spectra Chimica Acta A]
  {{10.1016/0584-8539(94)E0045-C}}, {51}, 39

\bibitem[\protect\citeauthoryear{Anacona, Bogey, Davies, Demuynck  \&
  Destombes}{Anacona et~al.}{1986}]{86AnBoDa.NS}
Anacona J.~R.,  Bogey M.,  Davies P.~B.,  Demuynck C.,   Destombes J.~L.,
  {1986}, \mn@doi [Mol. Phys.] {{10.1080/00268978600101921}}, {59}, 81

\bibitem[\protect\citeauthoryear{Azzam, Yurchenko, Tennyson  \& Naumenko}{Azzam
  et~al.}{2016}]{jt640}
Azzam A. A.~A.,  Yurchenko S.~N.,  Tennyson J.,   Naumenko O.~V.,  2016,
  \mn@doi [MNRAS] {10.1093/mnras/stw1133}, 460, 4063

\bibitem[\protect\citeauthoryear{Baeck \& Lee}{Baeck \&
  Lee}{1990}]{90BaLexx.SH}
Baeck K.~K.,  Lee Y.~S.,  1990, \mn@doi [J. Chem. Phys.] {10.1063/1.459572},
  93, 5775

\bibitem[\protect\citeauthoryear{Barklem \& Collet}{Barklem \&
  Collet}{2016}]{16BaCoxx.partfunc}
Barklem P.~S.,  Collet R.,  2016, \mn@doi [A\&A] {10.1051/0004-6361/201526961},
  588, A96

\bibitem[\protect\citeauthoryear{{Belloche}, {M{\"u}ller}, {Menten}, {Schilke}
  \& {Comito}}{{Belloche} et~al.}{2013}]{13BeMuMe.NS}
{Belloche} A.,  {M{\"u}ller} H.~S.~P.,  {Menten} K.~M.,  {Schilke} P.,
  {Comito} C.,  2013, \mn@doi [A\&A] {10.1051/0004-6361/201321096}, 559, A47

\bibitem[\protect\citeauthoryear{Benidar, Farrenq, Guelachvili  \&
  Chackerian}{Benidar et~al.}{1991}]{91BeFaGu.SH}
Benidar A.,  Farrenq R.,  Guelachvili G.,   Chackerian C.,  1991, \mn@doi [J.
  Mol. Spectrosc.] {10.1016/0022-2852(91)90063-G}, 147, 383

\bibitem[\protect\citeauthoryear{Berdyugina \& Livingston}{Berdyugina \&
  Livingston}{2002}]{02BeLiXX.SH}
Berdyugina S.~V.,  Livingston W.~C.,  2002, \mn@doi [A\&A]
  {10.1051/0004-6361:20020364}, 387, L6

\bibitem[\protect\citeauthoryear{Bernath, Amano  \& Wong}{Bernath
  et~al.}{1983}]{83BeAmWo.SH}
Bernath P.~F.,  Amano T.,   Wong M.,  1983, \mn@doi [J. Mol. Spectrosc.]
  {10.1016/0022-2852(83)90199-6}, 98, 20

\bibitem[\protect\citeauthoryear{Bialski \& Grein}{Bialski \&
  Grein}{1976}]{76BiGrXX.NS}
Bialski M.,  Grein F.,  1976, \mn@doi [J. Mol. Spectrosc.]
  {10.1016/0022-2852(76)90322-2}, 61, 321

\bibitem[\protect\citeauthoryear{{Biver}}{{Biver}}{2005}]{05Biver.NS}
{Biver} N.,  2005, in {Wilson} A.,  ed.,  ESA Special Publication Vol. 577, ESA
  Special Publication. pp 151--156

\bibitem[\protect\citeauthoryear{Brites, Hammoutene  \& Hochlaf}{Brites
  et~al.}{2008}]{08BrHaHo.SH}
Brites V.,  Hammoutene D.,   Hochlaf M.,  2008, \mn@doi [J. Phys.B: At. Mol.
  Phys.] {10.1088/0953-4075/41/4/045101}, 41, 045101

\bibitem[\protect\citeauthoryear{Brown \& Merer}{Brown \&
  Merer}{1979}]{79BrMexx.methods}
Brown J.~M.,  Merer A.~J.,  1979, \mn@doi [J. Mol. Spectrosc.]
  {10.1016/0022-2852(79)90172-3}, 74, 488

\bibitem[\protect\citeauthoryear{Bruna \& Hirsch}{Bruna \&
  Hirsch}{1987}]{87BrHixx.SH}
Bruna P.~J.,  Hirsch G.,  1987, \mn@doi [Mol. Phys.]
  {10.1080/00268978700101851}, 61, 1359

\bibitem[\protect\citeauthoryear{{Canaves}, {de Almeida}, {Boice}  \&
  {Sanzovo}}{{Canaves} et~al.}{2002}]{02CaAlBo.NS}
{Canaves} M.~V.,  {de Almeida} A.~A.,  {Boice} D.~C.,   {Sanzovo} G.~C.,  2002,
  \mn@doi [Earth Moon and Planets] {10.1023/A:1021582300423}, 90, 335

\bibitem[\protect\citeauthoryear{{Canaves}, {de Almeida}, {Boice}  \&
  {Sanzovo}}{{Canaves} et~al.}{2007}]{07CaAlBo.NS}
{Canaves} M.~V.,  {de Almeida} A.~A.,  {Boice} D.~C.,   {Sanzovo} G.~C.,  2007,
  \mn@doi [{Adv. Space Res.}] {10.1016/j.asr.2006.09.040}, 39, 451

\bibitem[\protect\citeauthoryear{{Carrington}, {Howard}, {Levy}  \&
  {Robertson}}{{Carrington} et~al.}{1968}]{68CaHoLe.NS}
{Carrington} A.,  {Howard} B.~J.,  {Levy} D.~H.,   {Robertson} J.~C.,  1968,
  \mn@doi [Mol. Phys.] {10.1080/00268976800100961}, 15, 187

\bibitem[\protect\citeauthoryear{{Charnley}}{{Charnley}}{1997}]{97Charnl.NS}
{Charnley} S.~B.,  1997, \mn@doi [ApJ] {10.1086/304011}, 481, 396

\bibitem[\protect\citeauthoryear{{Continetti}, {Balko}  \& {Lee}}{{Continetti}
  et~al.}{1991}]{91CoBaLe.SH}
{Continetti} R.~E.,  {Balko} B.~A.,   {Lee} Y.~T.,  1991, \mn@doi [Chem. Phys.
  Lett.] {10.1016/0009-2614(91)90097-S}, 182, 400

\bibitem[\protect\citeauthoryear{Csaszar, Leininger  \& Burcat}{Csaszar
  et~al.}{2003}]{03CsLeBu.SH}
Csaszar A.~G.,  Leininger M.~L.,   Burcat A.,  2003, \mn@doi [J. Phys. Chem. A]
  {10.1021/jp026605u}, 107, 2061

\bibitem[\protect\citeauthoryear{{Czernek} \& {{\v Z}ivn{\'y}}}{{Czernek} \&
  {{\v Z}ivn{\'y}}}{2004}]{04CzViXX.NS}
{Czernek} J.,  {{\v Z}ivn{\'y}} O.,  2004, \mn@doi [Chem. Phys.]
  {10.1016/j.chemphys.2004.05.014}, 303, 137

\bibitem[\protect\citeauthoryear{{Duley}, {Millar}  \& {Williams}}{{Duley}
  et~al.}{1980}]{80DuMiWi.SH}
{Duley} W.~W.,  {Millar} T.~J.,   {Williams} D.~A.,  1980, \mn@doi [MNRAS]
  {10.1093/mnras/192.4.945}, 192, 945

\bibitem[\protect\citeauthoryear{{Dunning Jr.}}{{Dunning
  Jr.}}{1989}]{89Dunnin.ai}
{Dunning Jr.} T.~H.,  1989, \mn@doi [J. Chem. Phys.] {10.1063/1.456153}, 90,
  1007

\bibitem[\protect\citeauthoryear{{Eliet}, {Martin-Drumel}, {Guinet}, {Hindle},
  {Mouret}, {Bocquet}  \& {Cuisset}}{{Eliet} et~al.}{2011}]{11ElMaGu.SH}
{Eliet} S.,  {Martin-Drumel} M.-A.,  {Guinet} M.,  {Hindle} F.,  {Mouret} G.,
  {Bocquet} R.,   {Cuisset} A.,  2011, \mn@doi [J. Mol. Spectrosc.]
  {10.1016/j.molstruc.2011.07.055}, 1006, 13

\bibitem[\protect\citeauthoryear{Furtenbacher, {Cs\'asz\'ar}  \&
  Tennyson}{Furtenbacher et~al.}{2007}]{MARVEL}
Furtenbacher T.,  {Cs\'asz\'ar} A.~G.,   Tennyson J.,  2007, J. Mol.
  Spectrosc., 245, 115

\bibitem[\protect\citeauthoryear{Gao, Gao  \& Gong}{Gao
  et~al.}{2013}]{13GaGaGo.NS}
Gao Y.,  Gao T.,   Gong M.,  2013, \mn@doi [J. Quant. Spectrosc. Radiat.
  Transf.] {10.1016/j.jqsrt.2013.06.014}, 129, 193

\bibitem[\protect\citeauthoryear{{Glockler} \& {Horwitz}}{{Glockler} \&
  {Horwitz}}{1939}]{39GlHoXX.SH}
{Glockler} G.,  {Horwitz} W.,  1939, \mn@doi [J. Chem. Phys.]
  {10.1063/1.1750544}, 7, 857

\bibitem[\protect\citeauthoryear{Gottlieb, Ball, Gottlieb, Lada  \&
  Penfield}{Gottlieb et~al.}{1975}]{75GoBaGo.NS}
Gottlieb C.~A.,  Ball J.~A.,  Gottlieb E.~W.,  Lada C.~J.,   Penfield H.,
  1975, ApJ, 200, L147

\bibitem[\protect\citeauthoryear{{Heiles} \& {Turner}}{{Heiles} \&
  {Turner}}{1971}]{71HeTuXX.SH}
{Heiles} C.~E.,  {Turner} B.~E.,  1971, ApJL, 8, 89

\bibitem[\protect\citeauthoryear{Hirst \& Guest}{Hirst \&
  Guest}{1982}]{82HiGuxx.SH}
Hirst D.~M.,  Guest M.~F.,  1982, \mn@doi [Mol. Phys.]
  {10.1080/00268978200101301}, 46, 427

\bibitem[\protect\citeauthoryear{{Irvine}, {Lovell}, {Senay}, {Matthews},
  {Metz}, {Meier}  \& {McGonagle}}{{Irvine} et~al.}{1999}]{99IrLoSe.NS}
{Irvine} W.~M.,  {Lovell} A.~J.,  {Senay} M.,  {Matthews} H.~E.,  {Metz} R.~B.,
   {Meier} R.,   {McGonagle} D.,  1999, in AAS/Division for Planetary Sciences
  Meeting Abstracts \#31. p. 32.03

\bibitem[\protect\citeauthoryear{Karna \& Grein}{Karna \&
  Grein}{1986}]{86KaGrXX.NS}
Karna S.~P.,  Grein F.,  1986, \mn@doi [J. Mol. Spectrosc.]
  {10.1016/0022-2852(86)90004-4}, 120, 284

\bibitem[\protect\citeauthoryear{{Karpfen}, {Schuster}, {Petkov}  \&
  {Lischka}}{{Karpfen} et~al.}{1978}]{78KaScPe.NS}
{Karpfen} A.,  {Schuster} P.,  {Petkov} J.,   {Lischka} H.,  1978, \mn@doi [J.
  Chem. Phys.] {10.1063/1.436196}, 68, 3884

\bibitem[\protect\citeauthoryear{{Kashinski}, {Talbi}, {Hickman}, {Di Nallo},
  {Colboc}, {Chakrabarti}, {Schneider}  \& {Mezei}}{{Kashinski}
  et~al.}{2017}]{17KaTaHi.SH}
{Kashinski} D.~O.,  {Talbi} D.,  {Hickman} A.~P.,  {Di Nallo} O.~E.,  {Colboc}
  F.,  {Chakrabarti} K.,  {Schneider} I.~F.,   {Mezei} J.~Z.,  2017, \mn@doi
  [J. Chem. Phys.] {10.1063/1.4983690}, 146, 204109

\bibitem[\protect\citeauthoryear{{Krishna Swamy} \& {Wallis}}{{Krishna Swamy}
  \& {Wallis}}{1987}]{87SwWaXX.SH}
{Krishna Swamy} K.~S.,  {Wallis} M.~K.,  1987, \mn@doi [MNRAS]
  {10.1093/mnras/228.2.305}, 228, 305

\bibitem[\protect\citeauthoryear{{Krishna Swamy} \& {Wallis}}{{Krishna Swamy}
  \& {Wallis}}{1988}]{88KrWaXX.SH}
{Krishna Swamy} K.~S.,  {Wallis} M.~K.,  1988, A\&AS, 74, 227

\bibitem[\protect\citeauthoryear{Lee, Seto, Hirao, Bernath  \& Le~Roy}{Lee
  et~al.}{1999}]{EMO}
Lee E.~G.,  Seto J.~Y.,  Hirao T.,  Bernath P.~F.,   Le~Roy R.~J.,  1999,
  \mn@doi [J. Mol. Spectrosc.] {10.1006/jmsp.1998.7789}, 194, 197

\bibitem[\protect\citeauthoryear{Lee, Ozeki  \& Saito}{Lee
  et~al.}{1995}]{95LeOzSa.NS}
Lee S.~K.,  Ozeki H.,   Saito S.,  {1995}, \mn@doi [ApJS] {{10.1086/192165}},
  {98}, 351

\bibitem[\protect\citeauthoryear{{Leurini} et~al.,}{{Leurini}
  et~al.}{2006}]{06LeRoTh.NS}
{Leurini} S.,  et~al., 2006, \mn@doi [A\&A] {10.1051/0004-6361:20065555}, 454,
  L47

\bibitem[\protect\citeauthoryear{{Lewis} \& {White}}{{Lewis} \&
  {White}}{1939}]{39LeWhXX.SH}
{Lewis} M.~N.,  {White} J.~U.,  1939, \mn@doi [Phys. Rev.]
  {10.1103/PhysRev.55.894}, 55, 894

\bibitem[\protect\citeauthoryear{Lie, Peyerimhoff  \& Buenker}{Lie
  et~al.}{1985}]{85LiPeBu.NS}
Lie G.~C.,  Peyerimhoff S.~D.,   Buenker R.~J.,  1985, \mn@doi [J. Chem. Phys.]
  {10.1063/1.448264}, 82, 2672

\bibitem[\protect\citeauthoryear{Lodi, Yurchenko  \& Tennyson}{Lodi
  et~al.}{2015}]{jt599}
Lodi L.,  Yurchenko S.~N.,   Tennyson J.,  2015, \mn@doi [Mol. Phys.]
  {10.1080/00268976.2015.1029996}, 113, 1559

\bibitem[\protect\citeauthoryear{Lovas \& Suenram}{Lovas \&
  Suenram}{1982}]{82LoSuXX.NS}
Lovas F.~J.,  Suenram R.~D.,  1982, \mn@doi [J. Mol. Spectrosc.]
  {10.1016/0022-2852(82)90177-1}, 93, 416

\bibitem[\protect\citeauthoryear{{Lovas}, {Johnson}  \& {Snyder}}{{Lovas}
  et~al.}{1979}]{79LoJoSn.NS}
{Lovas} F.~J.,  {Johnson} D.~R.,   {Snyder} L.~E.,  1979, \mn@doi [ApJS]
  {10.1086/190626}, 41, 451

\bibitem[\protect\citeauthoryear{{Mart{\'{\i}}n}}{{Mart{\'{\i}}n}}{2005}]{05Ma%
rtin.NS}
{Mart{\'{\i}}n} S.,  2005, in {H{\"u}ttmeister} S.,  {Manthey} E.,  {Bomans}
  D.,   {Weis} K.,  eds,  AIP Conf. Ser. Vol. 783, The Evolution of Starbursts.
  pp 148--154, \mn@doi{10.1063/1.2034979}

\bibitem[\protect\citeauthoryear{Martin-Drumel, Eliet, Pirali, Guinet, Hindle,
  Mouret  \& Cuisset}{Martin-Drumel et~al.}{2012}]{12MaElPi.SH}
Martin-Drumel M.~A.,  Eliet S.,  Pirali O.,  Guinet M.,  Hindle F.,  Mouret G.,
    Cuisset A.,  2012, \mn@doi [Chem. Phys. Lett.]
  {10.1016/j.cplett.2012.08.027}, 550, 8

\bibitem[\protect\citeauthoryear{{Mart{\'{\i}}n}, {Mauersberger},
  {Mart{\'{\i}}n-Pintado}, {Garc{\'{\i}}a-Burillo}  \&
  {Henkel}}{{Mart{\'{\i}}n} et~al.}{2003}]{03MaMaMa.NS}
{Mart{\'{\i}}n} S.,  {Mauersberger} R.,  {Mart{\'{\i}}n-Pintado} J.,
  {Garc{\'{\i}}a-Burillo} S.,   {Henkel} C.,  2003, \mn@doi [A\&A]
  {10.1051/0004-6361:20031442}, 411, L465

\bibitem[\protect\citeauthoryear{{Mart{\'{\i}}n}, {Mart{\'{\i}}n-Pintado},
  {Mauersberger}, {Henkel}  \& {Garc{\'{\i}}a-Burillo}}{{Mart{\'{\i}}n}
  et~al.}{2005}]{05MaMaMa.NS}
{Mart{\'{\i}}n} S.,  {Mart{\'{\i}}n-Pintado} J.,  {Mauersberger} R.,  {Henkel}
  C.,   {Garc{\'{\i}}a-Burillo} S.,  2005, \mn@doi [ApJ] {10.1086/426888}, 620,
  210

\bibitem[\protect\citeauthoryear{Matsumura, Kawaguchi, Nagai, Yamada  \&
  Hirota}{Matsumura et~al.}{1980}]{80MaKaNa.NS}
Matsumura K.,  Kawaguchi K.,  Nagai K.,  Yamada C.,   Hirota E.,  1980, \mn@doi
  [J. Mol. Spectrosc.] {10.1016/0022-2852(80)90239-8}, 84, 68

\bibitem[\protect\citeauthoryear{McCoy}{McCoy}{1998}]{98Mcxxxx.SH}
McCoy A.~B.,  1998, \mn@doi [J. Chem. Phys.] {10.1063/1.476546}, 109, 170

\bibitem[\protect\citeauthoryear{{McGonagle}, {Irvine}  \& {Minh}}{{McGonagle}
  et~al.}{1992}]{92McIrMi.NS}
{McGonagle} D.,  {Irvine} W.,   {Minh} Y.,  1992, in {Singh} P.~D.,  ed.,  IAU
  Symposium Vol. 150, Astrochemistry of Cosmic Phenomena. p.~227

\bibitem[\protect\citeauthoryear{{McGonagle}, {Irvine}  \&
  {Ohishi}}{{McGonagle} et~al.}{1994}]{94McIrOh.NS}
{McGonagle} D.,  {Irvine} W.~M.,   {Ohishi} M.,  1994, \mn@doi [ApJ]
  {10.1086/173755}, 422, 621

\bibitem[\protect\citeauthoryear{McKemmish, Yurchenko  \& Tennyson}{McKemmish
  et~al.}{2016}]{jt644}
McKemmish L.~K.,  Yurchenko S.~N.,   Tennyson J.,  2016, \mn@doi [MNRAS]
  {10.1093/mnras/stw1969}, 463, 771

\bibitem[\protect\citeauthoryear{Medvedev, Meshkov, Stolyarov, Ushakov  \&
  Gordon}{Medvedev et~al.}{2016}]{16MeMeSt}
Medvedev E.~S.,  Meshkov V.~V.,  Stolyarov A.~V.,  Ushakov V.~G.,   Gordon
  I.~E.,  2016, \mn@doi [J. Mol. Spectrosc.]
  {http://dx.doi.org/10.1016/j.jms.2016.06.013}, 330, 36

\bibitem[\protect\citeauthoryear{{Meeks}, {Gordon}  \& {Litvak}}{{Meeks}
  et~al.}{1969}]{69MeGoLi.SH}
{Meeks} M.~L.,  {Gordon} M.~A.,   {Litvak} M.~M.,  1969, \mn@doi [Science]
  {10.1126/science.163.3863.173}, 163, 173

\bibitem[\protect\citeauthoryear{Meerts \& Dymanus}{Meerts \&
  Dymanus}{1974}]{74MeDyxx.SH}
Meerts W.~L.,  Dymanus A.,  1974, \mn@doi [ApJ] {10.1086/181389}, 187, L45

\bibitem[\protect\citeauthoryear{{Meier} et~al.,}{{Meier}
  et~al.}{2015}]{15MeWaBo.NS}
{Meier} D.~S.,  et~al., 2015, \mn@doi [ApJ] {10.1088/0004-637X/801/1/63}, 801,
  63

\bibitem[\protect\citeauthoryear{Meuwly \& Hutson}{Meuwly \&
  Hutson}{1999}]{99MeHuxx.methods}
Meuwly M.,  Hutson J.~M.,  1999, \mn@doi [J. Chem. Phys.] {{10.1063/1.478744}},
  {110}, 8338

\bibitem[\protect\citeauthoryear{{M\"uller}, {Schl\"oder}, Stutzki  \&
  Winnewisser}{{M\"uller} et~al.}{2005}]{CDMS}
{M\"uller} H. S.~P.,  {Schl\"oder} F.,  Stutzki J.,   Winnewisser G.,  2005, J.
  Molec. Struct. (THEOCHEM), 742, 215

\bibitem[\protect\citeauthoryear{{Narasimham} \&
  {Balasubramanian}}{{Narasimham} \& {Balasubramanian}}{1971}]{71NaBaXX.NS}
{Narasimham} N.~A.,  {Balasubramanian} T.~K.,  1971, \mn@doi [J. Mol.
  Spectrosc.] {10.1016/0022-2852(71)90253-0}, 40, 511

\bibitem[\protect\citeauthoryear{{Neufeld} et~al.,}{{Neufeld}
  et~al.}{2012}]{12NeFaGe.SH}
{Neufeld} D.~A.,  et~al., 2012, \mn@doi [A\&A] {10.1051/0004-6361/201218870},
  542, L6

\bibitem[\protect\citeauthoryear{{Oppenheimer} \& {Dalgarno}}{{Oppenheimer} \&
  {Dalgarno}}{1974}]{74OpDaXX.NS}
{Oppenheimer} M.,  {Dalgarno} A.,  1974, \mn@doi [ApJ] {10.1086/152618}, 187,
  231

\bibitem[\protect\citeauthoryear{Patrascu, Tennyson  \& Yurchenko}{Patrascu
  et~al.}{2015}]{jt598}
Patrascu A.~T.,  Tennyson J.,   Yurchenko S.~N.,  2015, \mn@doi [MNRAS]
  {10.1093/mnras/stv507}, 449, 3613

\bibitem[\protect\citeauthoryear{Paulose, Barton, Yurchenko  \&
  Tennyson}{Paulose et~al.}{2015}]{jt615}
Paulose G.,  Barton E.~J.,  Yurchenko S.~N.,   Tennyson J.,  2015, \mn@doi
  [MNRAS] {10.1093/mnras/stv1543}, 454, 1931

\bibitem[\protect\citeauthoryear{Peterson \& {Dunning Jr.}}{Peterson \&
  {Dunning Jr.}}{2002}]{02PeDuxx.ai}
Peterson K.~A.,  {Dunning Jr.} T.~H.,  2002, \mn@doi [J. Chem. Phys.]
  {10.1063/1.1520138}, 117, 10548

\bibitem[\protect\citeauthoryear{Prajapat, Jagoda, Lodi, Gorman, Yurchenko  \&
  Tennyson}{Prajapat et~al.}{2017}]{jt703}
Prajapat L.,  Jagoda P.,  Lodi L.,  Gorman M.~N.,  Yurchenko S.~N.,   Tennyson
  J.,  2017, \mn@doi [MNRAS] {10.1093/mnras/stx2229}, 472, 3648

\bibitem[\protect\citeauthoryear{Qui-Xia, Tao  \& Yun-Guang}{Qui-Xia
  et~al.}{2008}]{08LiGaZh.SH}
Qui-Xia L.,  Tao G.,   Yun-Guang Z.,  2008, Chin. Phys. B, 17, 2040

\bibitem[\protect\citeauthoryear{Raimondi, Tantardini  \& Simonetta}{Raimondi
  et~al.}{1975}]{75RaTaSi.SH}
Raimondi M.,  Tantardini G.~F.,   Simonetta M.,  1975, \mn@doi [Mol. Phys.]
  {10.1080/00268977500102271}, 30, 703

\bibitem[\protect\citeauthoryear{Ram, Bernath, Engleman  \& Brault}{Ram
  et~al.}{1995}]{95RaBeEn.SH}
Ram R.~S.,  Bernath P.~F.,  Engleman R.,   Brault J.~W.,  1995, \mn@doi [J.
  Mol. Spectrosc.] {10.1006/jmsp.1995.1153}, 172, 34

\bibitem[\protect\citeauthoryear{{Ravichandran}, {Williams}  \&
  {Fletcher}}{{Ravichandran} et~al.}{1994}]{94RaWiFl.SH}
{Ravichandran} K.,  {Williams} R.,   {Fletcher} T.~R.,  1994, \mn@doi [Chem.
  Phys. Lett.] {10.1016/0009-2614(93)E1411-9}, 217, 375

\bibitem[\protect\citeauthoryear{Resende \& Ornellas}{Resende \&
  Ornellas}{2001}]{01ReOrx1.SH}
Resende S.~M.,  Ornellas F.~R.,  2001, \mn@doi [J. Chem. Phys.]
  {10.1063/1.1381577}, 115, 2178

\bibitem[\protect\citeauthoryear{{Rodgers} \& {Charnley}}{{Rodgers} \&
  {Charnley}}{2006}]{06RoChXX.NS}
{Rodgers} S.~D.,  {Charnley} S.~B.,  2006, \mn@doi [Adv. Space Res.]
  {10.1016/j.asr.2005.10.006}, 38, 1928

\bibitem[\protect\citeauthoryear{{Salahub} \& {Messmer}}{{Salahub} \&
  {Messmer}}{1976}]{76SaMeXX.NS}
{Salahub} D.~R.,  {Messmer} R.~P.,  1976, \mn@doi [J. Chem. Phys.]
  {10.1063/1.432480}, 64, 2039

\bibitem[\protect\citeauthoryear{Sauval \& Tatum}{Sauval \&
  Tatum}{1984}]{84SaTaxx.partfunc}
Sauval A.~J.,  Tatum J.~B.,  1984, ApJS, 56, 193

\bibitem[\protect\citeauthoryear{Semenov, Yurchenko  \& Tennyson}{Semenov
  et~al.}{2017}]{jt656}
Semenov M.,  Yurchenko S.~N.,   Tennyson J.,  2017, \mn@doi [J. Mol.
  Spectrosc.] {10.1016/j.jms.2016.11.004}, 330, 57

\bibitem[\protect\citeauthoryear{{Shi}, {Xing}, {Sun}  \& {Zhu}}{{Shi}
  et~al.}{2012}]{12ShXiSu.NS}
{Shi} D.~H.,  {Xing} W.,  {Sun} J.~F.,   {Zhu} Z.~L.,  2012, \mn@doi [Eur.
  Phys. J. D] {10.1140/epjd/e2012-30206-2}, 66, 173

\bibitem[\protect\citeauthoryear{Sinha, Burkholder, Hammer  \& Howard}{Sinha
  et~al.}{1988}]{88SiBuHa.NS}
Sinha A.,  Burkholder J.~B.,  Hammer P.~D.,   Howard C.~J.,  {1988}, \mn@doi
  [J. Mol. Spectrosc.] {{10.1016/0022-2852(88)90093-8}}, {130}, 466

\bibitem[\protect\citeauthoryear{Skokov, Peterson  \& Bowman}{Skokov
  et~al.}{1999}]{99SkPeBo.methods}
Skokov S.,  Peterson K.~A.,   Bowman J.~M.,  1999, \mn@doi [Chem. Phys. Lett.]
  {{10.1016/S0009-2614(99)00996-3}}, {312}, 494

\bibitem[\protect\citeauthoryear{{Somerville}}{{Somerville}}{1977}]{77Somerv.N%
S}
{Somerville} W.~B.,  1977, \mn@doi [{Rep. Prog. Phys}]
  {10.1088/0034-4885/40/5/001}, 40, 483

\bibitem[\protect\citeauthoryear{Szalay, M{\"u}ller, Gidofalvi, Lischka  \&
  Shepard}{Szalay et~al.}{2012}]{12SzMuGi.ai}
Szalay P.,  M{\"u}ller T.,  Gidofalvi G.,  Lischka H.,   Shepard R.,  2012,
  Chem. Rev., 112, 108

\bibitem[\protect\citeauthoryear{Tennyson, Hulme, Naim  \& Yurchenko}{Tennyson
  et~al.}{2016a}]{jt624}
Tennyson J.,  Hulme K.,  Naim O.~K.,   Yurchenko S.~N.,  2016a, \mn@doi [J.
  Phys. B: At. Mol. Opt. Phys.] {10.1088/0953-4075/49/4/044002}, 49, 044002

\bibitem[\protect\citeauthoryear{Tennyson, Lodi, McKemmish  \&
  Yurchenko}{Tennyson et~al.}{2016b}]{jt632}
Tennyson J.,  Lodi L.,  McKemmish L.~K.,   Yurchenko S.~N.,  2016b, J. Phys. B:
  At. Mol. Opt. Phys., 49, 102001

\bibitem[\protect\citeauthoryear{Tennyson et~al.,}{Tennyson
  et~al.}{2016c}]{jt631}
Tennyson J.,  et~al., 2016c, \mn@doi [J. Mol. Spectrosc.]
  {10.1016/j.jms.2016.05.002}, 327, 73

\bibitem[\protect\citeauthoryear{Uehara \& Morino}{Uehara \&
  Morino}{1969}]{69UeMoXX.NS}
Uehara H.,  Morino Y.,  1969, \mn@doi [Mol. Phys.] {10.1080/00268976900100991},
  17, 239

\bibitem[\protect\citeauthoryear{Underwood, Tennyson, Yurchenko, Huang,
  Schwenke, Lee, Clausen  \& Fateev}{Underwood et~al.}{2016a}]{jt635}
Underwood D.~S.,  Tennyson J.,  Yurchenko S.~N.,  Huang X.,  Schwenke D.~W.,
  Lee T.~J.,  Clausen S.,   Fateev A.,  2016a, \mn@doi [MNRAS]
  {10.1093/mnras/stw849}, 459, 3890

\bibitem[\protect\citeauthoryear{Underwood, Tennyson, Yurchenko, Clausen  \&
  Fateev}{Underwood et~al.}{2016b}]{jt641}
Underwood D.~S.,  Tennyson J.,  Yurchenko S.~N.,  Clausen S.,   Fateev A.,
  2016b, \mn@doi [MNRAS] {10.1093/mnras/stw1828}, 462, 4300

\bibitem[\protect\citeauthoryear{Upadhyay, Conway, Tennyson  \&
  Yurchenko}{Upadhyay et~al.}{2018}]{jt724}
Upadhyay A.,  Conway E.~K.,  Tennyson J.,   Yurchenko S.~N.,  2018, MNRAS

\bibitem[\protect\citeauthoryear{Vamhindi \& Nsangou}{Vamhindi \&
  Nsangou}{2016}]{16VsNsxx.SH}
Vamhindi B. S. D.~R.,  Nsangou M.,  2016, \mn@doi [Mol. Phys.]
  {10.1080/00268976.2016.1191690}, 114, 2204

\bibitem[\protect\citeauthoryear{Vidal, Loison, Jaziri, Ruaud, Gratier  \&
  Wakelam}{Vidal et~al.}{2017}]{17ViLoJa.SH}
Vidal T. H.~G.,  Loison J.-C.,  Jaziri A.~Y.,  Ruaud M.,  Gratier P.,   Wakelam
  V.,  {2017}, \mn@doi [MNRAS] {{10.1093/mnras/stx828}}, {469}, 435

\bibitem[\protect\citeauthoryear{Vidler \& Tennyson}{Vidler \&
  Tennyson}{2000}]{jt263}
Vidler M.,  Tennyson J.,  2000, J. Chem. Phys., 113, 9766

\bibitem[\protect\citeauthoryear{{Visscher}, {Lodders}  \& {Fegley}}{{Visscher}
  et~al.}{2006}]{06ViLoFe.SH}
{Visscher} C.,  {Lodders} K.,   {Fegley} Jr. B.,  2006, \mn@doi [ApJ]
  {10.1086/506245}, 648, 1181

\bibitem[\protect\citeauthoryear{Werner, Knowles, Knizia, Manby  \&
  Sch\"utz}{Werner et~al.}{2012}]{MOLPRO}
Werner H.-J.,  Knowles P.~J.,  Knizia G.,  Manby F.~R.,   Sch\"utz M.,  2012,
  \mn@doi [WIREs Comput. Mol. Sci.] {10.1002/wcms.82}, 2, 242

\bibitem[\protect\citeauthoryear{Western}{Western}{2017}]{PGOPHER}
Western C.~M.,  2017, \mn@doi [J. Quant. Spectrosc. Radiat. Transf.]
  {10.1016/j.jqsrt.2016.04.010}, 186, 221

\bibitem[\protect\citeauthoryear{Winkel \& Davis}{Winkel \&
  Davis}{1984}]{84WiDaxx.SH}
Winkel R.~J.,  Davis S.~P.,  1984, \mn@doi [Can. J. Phys.] {10.1139/p84-189},
  62, 1420

\bibitem[\protect\citeauthoryear{Wong, Yurchenko, Bernath, Mueller, McConkey
  \& Tennyson}{Wong et~al.}{2017}]{jt686}
Wong A.,  Yurchenko S.~N.,  Bernath P.,  Mueller H. S.~P.,  McConkey S.,
  Tennyson J.,  2017, \mn@doi [MNRAS] {10.1093/mnras/stx1211}, 470, 882

\bibitem[\protect\citeauthoryear{{Woods}, {Occhiogrosso}, {Viti}, {Ka{\v
  n}uchov{\'a}}, {Palumbo}  \& {Price}}{{Woods} et~al.}{2015}]{15WoOcVi.NS}
{Woods} P.~M.,  {Occhiogrosso} A.,  {Viti} S.,  {Ka{\v n}uchov{\'a}} Z.,
  {Palumbo} M.~E.,   {Price} S.~D.,  2015, \mn@doi [MNRAS]
  {10.1093/mnras/stv652}, 450, 1256

\bibitem[\protect\citeauthoryear{Woon \& Dunning}{Woon \&
  Dunning}{1993}]{93WoDuxx.ai}
Woon D.~E.,  Dunning T.~H.,  {1993}, J. Chem. Phys., {98}, 1358

\bibitem[\protect\citeauthoryear{Yamamura, Kawaguchi  \& Ridgway}{Yamamura
  et~al.}{2000}]{00YaKaRi.SH}
Yamamura I.,  Kawaguchi K.,   Ridgway S.~T.,  2000, \mn@doi [ApJ]
  {10.1086/312420}, 528, L33

\bibitem[\protect\citeauthoryear{Yurchenko, Lodi, Tennyson  \&
  Stolyarov}{Yurchenko et~al.}{2016a}]{jt609}
Yurchenko S.~N.,  Lodi L.,  Tennyson J.,   Stolyarov A.~V.,  2016a, \mn@doi
  [Comput. Phys. Commun.] {10.1016/j.cpc.2015.12.021}, 202, 262

\bibitem[\protect\citeauthoryear{Yurchenko, Blissett, Asari, Vasilios, Hill  \&
  Tennyson}{Yurchenko et~al.}{2016b}]{jt618}
Yurchenko S.~N.,  Blissett A.,  Asari U.,  Vasilios M.,  Hill C.,   Tennyson
  J.,  2016b, \mn@doi [MNRAS] {10.1093/mnras/stv2858}, 456, 4524

\bibitem[\protect\citeauthoryear{Yurchenko, Sinden, Lodi, Hill, Gorman  \&
  Tennyson}{Yurchenko et~al.}{2018}]{jt711}
Yurchenko S.~N.,  Sinden F.,  Lodi L.,  Hill C.,  Gorman M.~N.,   Tennyson J.,
  2018, \mn@doi [MNRAS] {10.1093/mnras/stx2738}, 473, 5324

\bibitem[\protect\citeauthoryear{Zahnle, Marley, Freedman, Lodders  \&
  Fortney}{Zahnle et~al.}{2009}]{09ZaMaFr.SH}
Zahnle K.,  Marley M.~S.,  Freedman R.~S.,  Lodders K.,   Fortney J.~J.,  2009,
  \mn@doi [Astrophys. J. Lett.] {10.1088/0004-637X/701/1/L20}, 701, L20

\bibitem[\protect\citeauthoryear{{Zhao}, {Galazutdinov}, {Linnartz}  \&
  {Kre{\l}owski}}{{Zhao} et~al.}{2015}]{15ZhGaLi.SH}
{Zhao} D.,  {Galazutdinov} G.~A.,  {Linnartz} H.,   {Kre{\l}owski} J.,  2015,
  \mn@doi [A\&A] {10.1051/0004-6361/201526488}, 579, L1

\makeatother
\end{thebibliography}

\label{lastpage}
\end{document}